%% file: main.tex
\pdfoutput=1
\documentclass[aps, prd, floats, floatfix, superscriptaddress, onecolumn,PRD,nofootinbib,amsmath,amssymb]{revtex4-2}

\usepackage[T1]{fontenc}
\usepackage[utf8]{inputenc}
\usepackage{lmodern}

\usepackage{MnSymbol}
\usepackage{amssymb}
\usepackage{amsfonts}
\usepackage{eufrak}

\usepackage[dvipsnames, usenames]{xcolor}
\usepackage{graphicx}
\usepackage{xspace}
\usepackage[normalem]{ulem}
\usepackage{bm}
\usepackage{slashed}
\usepackage{appendix}

\definecolor{linkcolor}{rgb}{0.0,0.3,0.5}
\usepackage[hypertexnames=false, unicode, colorlinks=true, linkcolor=linkcolor,
citecolor=linkcolor, filecolor=linkcolor,urlcolor=linkcolor,
pdfusetitle]{hyperref}
\usepackage[all]{hypcap}
\usepackage{microtype}

\usepackage[english]{babel}

\usepackage{import}

\usepackage{listings}
\definecolor{codegreen}{rgb}{0,0.6,0}
\definecolor{codegray}{rgb}{0.5,0.5,0.5}
\definecolor{codepurple}{rgb}{0.58,0,0.82}
\definecolor{backcolour}{rgb}{0.95,0.95,0.92}
\lstdefinestyle{mystyle}{
    backgroundcolor=\color{backcolour},   
    commentstyle=\color{codegreen},
    keywordstyle=\color{magenta},
    numberstyle=\tiny\color{codegray},
    stringstyle=\color{codepurple},
    basicstyle=\ttfamily,
    breakatwhitespace=false,         
    breaklines=true,                 
    captionpos=b,                    
    keepspaces=true,                 
    numbers=left,                    
    numbersep=5pt,                  
    showspaces=false,                
    showstringspaces=false,
    showtabs=false,                  
    tabsize=2
}
\renewcommand{\vec}[1] {\bm{#1}}

\newcommand{\RDMSM}{R_{\rm DM-SM}}
\newcommand{\RDMSMdark}{R_{\rm DM-SM,dark}}

\newcommand{\ratioDMSMtoDMDR}{\mathcal{R}}

\newcommand{\CLASSADM}{CLASS ADM\xspace}
\newcommand{\CLASSMAD}{MAD CLASS\xspace}

\begin{document}

\title{At the EDGES of MADness: phenomenology of relic millicharged atomic dark matter}

\newcommand\elehome{
\affiliation{Department of Physics, Ben-Gurion University of the Negev, Be'er Sheva 84105, Israel}
\affiliation{School of Physics and Astronomy, Tel-Aviv University, Tel-Aviv 69978, Israel}
}

\newcommand\elyhome{
\affiliation{Department of Physics, Ben-Gurion University of the Negev, Be'er Sheva 84105, Israel}
}

\newcommand\kimhome{
\affiliation{Texas Center for Cosmology and Astroparticle Physics, Weinberg Institute for Theoretical Physics, Department of Physics, The University of Texas at Austin, Austin, TX 78712, USA}
}

\author{Eleonora Vanzan}
\email{evanzan@tauex.tau.ac.il}
\elehome

\author{Kimberly K.~Boddy}
\kimhome

\author{Ely D. Kovetz}
\elyhome

\begin{abstract}
    Millicharged dark matter lives in a corner of parameter space that has been considerably shrinking over the past decades, due to constraints from cosmology, astrophysical systems, and ground-based experiments.
    A particularly interesting probe is the 21-cm absorption signal at Cosmic Dawn, which attracted interest in light of the EDGES claim: the reported excess absorption can be explained by millicharged dark matter with a fractional electric charge of~$10^{-6} \lesssim \epsilon \lesssim 10^{-4}$, but it must be a subpercent of the total dark matter abundance to remain consistent with the cosmic microwave background (CMB).
    We show that very low fractions of millicharged dark matter naturally arise from atomic dark matter: a dark sector model that replicates the physics of standard atomic hydrogen.
    Dark ions undergo the process of dark recombination, leaving a relic ionized component that, in the presence of kinetic mixing, is naturally millicharged.
    We modify the cosmological Boltzmann solver CLASS to incorporate this scenario, and we investigate the parameter space of millicharged atomic dark matter (MAD) that is allowed by CMB, large-scale structure, and supernovae data.
    We find there are cosmologically viable regions of MAD parameter space that are consistent with laboratory and astrophysical bounds and also overlap with the EDGES-defined window.
    Although the EDGES result is likely not a Cosmic Dawn signal, MAD leaves distinctive cosmological signatures, accessible to upcoming 21-cm observations.
\end{abstract}

\maketitle

\tableofcontents

\section{Introduction}

Millicharged dark matter (DM) is a dark-sector particle endowed with a fractional electric charge~$\epsilon e$, giving rise to rich cosmological phenomenology.
It interacts with (ionized) baryons through a momentum-transfer cross section with a Coulomb-like relative-velocity dependence~\cite{McDermott:2010pa,Boddy:2018wzy,Nguyen:2021cnb}, $\sigma(v)\propto v^{-4}$, typical of a massless or nearly massless mediator, resulting in a larger scattering rate at later times when velocities are lower.
DM-baryon scattering can impact the early Universe, modifying cosmic microwave background (CMB) anisotropies and suppressing small-scale structure, as well as dramatically affect late-time observables, e.g.,~around Cosmic Dawn and reionization by altering the evolution of the baryon temperature.
Even a small percentage of millicharged DM can provide an efficient baryon cooling channel around Cosmic Dawn and reionization~\cite{Munoz:2015bca,Munoz:2018pzp,Berlin:2018sjs,Barkana:2018qrx,Kovetz:2018zan,Liu:2019knx}, while leaving the early Universe, and in particular the CMB, nearly unaffected~\cite{Boddy:2018wzy,dePutter:2018xte}.
This possibility attracted renewed attention following the EDGES report of an anomalously deep 21-cm absorption feature at~$z \sim 17$ twice as deep as $\Lambda$CDM predictions~\cite{Bowman:2018yin}; although the result is controversial~\cite{Cang:2024gtp,Bevins:2022ajf}, it provides an interesting benchmark for current studies.
More generally, upcoming 21-cm line-intensity mapping surveys will provide powerful probes of any new physics that alters the thermal history of the intergalactic medium around these epochs~\cite{Munoz:2015bca}, motivating further investigation into the viability of millicharged models.

The region of parameter space in which millicharged DM explains the EDGES observation is limited.
CMB constraints require the millicharged component to comprise only a subpercent fraction of the total dark matter~\cite{Boddy:2018wzy,Kovetz:2018zan,dePutter:2018xte}, with further restrictions from the impact of long-range dark-sector interactions on structure formation~\cite{Ackerman:2008kmp,Cyr-Racine:2015ihg}.
Such a low interacting fraction, however, does not emerge naturally in simple, single-particle millicharged DM models~\cite{Creque-Sarbinowski:2019mcm,Berlin:2018sjs,Liu:2019knx}.
In order to produce observable 21-cm effects while remaining consistent with other bounds, the fraction of millicharged DM (in terms of the total DM abundance) must be~$\lesssim 0.4 \%$~\cite{Boddy:2018wzy,dePutter:2018xte} with fractional electric charge~$10^{-6} \lesssim \epsilon \lesssim 10^{-4}$~\cite{Kovetz:2018zan}.
In minimal models of millicharged DM, invoking such a low fractional abundance is ad hoc, which is not appealing from a theoretical standpoint.
It is therefore natural to ask whether there is a model in which such a small millicharged fraction arises automatically, as the output of the dark sector's own thermal history.

In this paper, we produce a small millicharged fraction in the context of millicharged atomic dark matter (MAD).
This scenario is based on a secluded sector of atomic dark matter (ADM)~\cite{Kaplan:2009de,Kaplan:2011yj,Cyr-Racine:2012tfp,Cyr-Racine:2013fsa,Cyr-Racine:2015ihg,Bansal:2022qbi,Essig:2018pzq}, which features two dark charged massive fermions (dark electron and dark proton) bound together by a massless dark~$U(1)$ photon.
ADM is a specific realization of a class of well-motivated high-energy theories, such as Twin Higgs models~\cite{Chacko:2018vss,Craig:2016lyx,GarciaGarcia:2015pnn}, rooted in solutions to the Hierarchy Problem, but it can also be viewed as a minimal realization of a more generic theory involving a hidden dark plasma in the early Universe, where the DM remains kinetically coupled to the dark radiation (DR)~\cite{Cyr-Racine:2015ihg}.
Previous studies of ADM focused on the cosmological signatures of dark radiation
and dark acoustic oscillations~\cite{Bansal:2022qbi,Cyr-Racine:2012tfp,Cyr-Racine:2013fsa,Cyr-Racine:2015ihg,Fan:2013tia,Fan:2013yva,Curtin:2020tkm}.
ADM mirrors the thermal history of the visible sector: at early times, the dark sector forms a tightly coupled plasma; as the Universe cools, dark recombination proceeds and the dark fermions bind into neutral dark hydrogen, leaving only a relic ionized fraction.

The key to obtaining an EDGES-like Cosmic Dawn signature is to introduce interactions with the visible sector.
We consider kinetic mixing between the dark and visible photons~\cite{Holdom:1985ag,Goldberg:1986nk,Feldman:2007wj}, so that the residual ionized component from dark recombination acquires a small millicharge~$\epsilon e$~\cite{Cline:2012is}.
We implement MAD into the cosmological Boltzmann solver CLASS v3.1.1~\cite{Blas:2011rf} and perform a systematic analysis of the CMB and matter power spectrum, treating the dark recombination history and DM-baryon scattering self-consistently.
Our work builds on a previous modification of CLASS for ADM~\cite{Bansal:2022qbi},\footnote{\href{https://github.com/jp-barron/class_adm-3.1}{\lstinline{https://github.com/jp-barron/class_adm-3.1}}} itself based on a mirror twin Higgs scenario~\cite{Bansal:2021dfh}.
We make our code \CLASSMAD publicly available.\footnote{\href{https://github.com/eleonoravanzan/mad_class}{\lstinline{https://github.com/eleonoravanzan/mad_class}}}

Our analysis yields two main findings.
First, MAD realizations populate the cosmologically viable mass-millicharge plane, with dark constituent masses~$\gtrsim 10 \, {\rm MeV}$ and millicharges in the open windows~$10^{-6} \lesssim \epsilon \lesssim 10^{-4}$, or~$\epsilon \lesssim 10^{-8}$, including a broad-compatibility band around masses that overlap with the parameter window invoked in EDGES-style explanations of baryon cooling at Cosmic Dawn.
Second, the interplay between the dark recombination history and the DM-baryon interactions opens regions of parameter space nearly invisible to the CMB and large-scale structure observables, yet it leaves distinctive imprints on the late-time baryon and dark baryon temperatures, impacting 21-cm observables.

The rest of this work is structured as follows.
We introduce the full model in Sec.~\ref{sec:model}, and the details of the DM-baryon interactions in Sec.~\ref{sec:interactions}.
Sec.~\ref{sec:thermal_history} describes the evolution of temperatures and cosmological perturbations, and introduces a convenient parameterization of DM-baryon interaction strength leveraging the thermal history of the dark sector.
Sec.~\ref{sec:analysis} describes the analysis, which is discussed in Sec.~\ref{sec:discussion}.
We conclude in Sec.~\ref{sec:conclusions}.

\section{The model}
\label{sec:model}

Our MAD model is a dark sector that consists of two fermionic particles, oppositely charged under a hidden~$U(1)_D$ gauge symmetry associated with a massless dark gauge boson (dark photon~$\gamma_D$) that kinetically mixes with the Standard Model (SM) photon.
Analogous to standard hydrogen, our fermions are the dark proton~$p_D$ and dark electron~$e_D$, and we assume~$m_{p_D} > m_{e_D}$.
Additionally, we assume there is a primordial asymmetry in the dark sector that results in a Universe with dark sector particles and not dark sector antiparticles.
The Lagrangian is
\begin{equation}
    \mathcal{L} = \Bar{p}_D \left( i \slashed{\Tilde{D}}_D - m_{p_D} \right) p_D + \Bar{e}_D \left( i \slashed{\Tilde{D}}_D - m_{e_D} \right) e_D -\frac{1}{4} (\Tilde{F}_D)_{\mu\nu} (\Tilde{F}_D)^{\mu\nu} -\frac{1}{4} \Tilde{F}_{\mu\nu} \Tilde{F}^{\mu\nu} -\frac{\sin\chi}{2} \Tilde{F}_{\mu\nu} (\Tilde{F}_D)^{\mu\nu} \, ,
\end{equation}
where~$\Tilde{D}_D^\mu = \partial^\mu + i g \Tilde{A}_D^\mu$, with~$g^2 = 4\pi \alpha_D$ the~$U(1)_D$ coupling constant, and~$\sin\chi$ parameterizes the amount of kinetic mixing.
The field strengths~$\Tilde{F}^{\mu\nu}$ and~$\Tilde{F}_D^{\mu\nu}$ are associated with the visible and dark gauge bosons~$\Tilde{A}^\mu$ and~$\Tilde{A}_D^\mu$, respectively.
We diagonalize the kinetic term via the field redefinitions~$\Tilde{A}^\mu = A^\mu/\cos\chi$ and~$\Tilde{A}_D^\mu = A_D^\mu -A^\mu \tan\chi$, which puts our Lagrangian in canonical form for the SM photon~$A^\mu$ and dark photon~$A_D^\mu$.
In this basis, the dark particles acquire a fractional electric charge~$\epsilon e$, with~$\epsilon = (g/e) \tan\chi$~\cite{Burrage:2009yz,Feldman:2007wj}.
If either~$\chi$ or~$g$ are small, the charge of the dark particles is also small,~$|\epsilon| \ll 1$.
The Lagrangian becomes
\begin{equation}
    \mathcal{L} = \Bar{p}_D \left( i \slashed{D}_D - m_{p_D} \right) p_D + \Bar{e}_D \left( i \slashed{D}_D - m_{e_D} \right) e_D - \frac{1}{4} (F_D)_{\mu\nu} (F_D)^{\mu\nu} -\frac{1}{4} F_{\mu\nu} F^{\mu\nu} + \epsilon e J_D^{\mu} A_{\mu} \, ,
\end{equation}
where~$J_D^{\mu} = \Bar{p}_D \gamma^{\mu} p_D - \Bar{e}_D \gamma^{\mu} e_D$.

At early times, the dark fermions and dark radiation form a tightly-coupled plasma in thermal equilibrium.\footnote{We restrict our parameter space to DM-SM interactions that are not strong enough to break the assumption of initial thermal equilibrium between DM and DR. Scenarios where an order unity of the dark matter content is tightly coupled to the baryons at early times are already excluded~\cite{Boddy:2018wzy,Cyr-Racine:2013fsa}.}
Dark hydrogen is able to form when the dark sector temperature drops below the dark hydrogen binding energy, initiating dark recombination: the dark fermions can combine into bound states, with some leftover dark ionization fraction~$x_D = n_{e_D,\rm free} / n_D$, where~$n_D$ is the number density of dark protons and dark hydrogen.
We do not consider dark nuclei, such as dark helium, which can arise if the dark proton is the bound state of a dark QCD~\cite{Bansal:2021dfh}.

We assume that dark ions and dark hydrogen together comprise the full abundance of dark matter. We only consider scenarios where dark recombination occurs prior to SM recombination to ensure that dark acoustic oscillations are confined to sufficiently small scales to satisfy CMB and LSS bounds~\cite{Bansal:2022qbi,Cyr-Racine:2013fsa}.
We also assume that DM particles are kept in thermal equilibrium by dark photon-dark fermion interactions at early times and by dark Coulomb scattering after dark decoupling.

Our model is described by five parameters,
\begin{equation}
    \left\{ \Delta N_D, \, \alpha_D, \, m_{p_D}, \, m_{e_D}, \, \epsilon \right\} \, .
\end{equation}
The first parameter,~$\Delta N_D$, is the contribution of DR to the effective number of relativistic neutrino-like species~$N_{\rm eff}$, with the important caveat that DR is freely streaming only after dark recombination.
We note that in generic ADM, the dark sector can be much colder than the visible one, for instance, if the two sectors are reheated asymmetrically after inflation~\cite{Berezhiani:1995am,Adshead:2016xxj}, or if they decoupled from each other at very early times so that the entropy released by subsequent SM annihilations heats only the visible bath~\cite{Feng:2008mu}.
In a MAD scenario, however, it is conceivable that the visible and dark sectors share a common reheating origin, in which case both baths start out at a common temperature~$T_{\rm rh}$ and evolve adiabatically and independently thereafter.
The comoving entropy of each sector is then separately conserved, fixing the temperature ratio at any later epoch,
\begin{equation}
\label{eq:xi}
    \xi = \frac{T_{\gamma,\rm dark}}{T_{\gamma}} = \left[ \frac{g_{*s}(T_{\gamma})}{g_{*s}(T_{\rm rh})} \, \frac{g_{*s,\rm dark}(T_{\rm rh})}{g_{*s,\rm dark}(T_{\gamma,\rm dark})} \right]^{1/3} \, ,
\end{equation}
where~$g_{*s}$ and~$g_{*s,\rm dark}$ denote the effective numbers of relativistic degrees of freedom entering the entropy densities of the visible and dark sectors, respectively.
By assuming that there are no other relativistic degrees of freedom in the dark sector, the temperature ratio can be linked to~$\Delta N_D$ via
\begin{equation}
    \Delta N_D = \frac{8}{7} \left( \frac{11}{4} \right)^{4/3} \xi^4 \simeq 4.4 \, \xi^4 \, .
\end{equation}
Since between reheating and recombination the SM bath loses far more relativistic species ($g_{*s}\!: 106.75 \to 3.91$) than the dark sector, where only the two dark fermions transfer their entropy to the dark photons ($g_{*s,\rm dark}\!: 9 \to 2$), the dark sector naturally ends up colder: for reheating above all mass thresholds, Eq.~\eqref{eq:xi} yields~$\xi \simeq 0.55$, corresponding to~$\Delta N_D \simeq 0.4$, allowed by BBN alone at the~$2\sigma$ level~\cite{Yeh:2022heq,Pitrou:2018cgg}.

The parameters~$\{ \alpha_D, \, m_{p_D}, \, m_{e_D} \}$ characterize the microphysical properties of ADM.
The binding energy of the dark hydrogen atoms is
\begin{equation}
    B_D = \frac{\mu_D \alpha_D^2}{2} \, ,
\end{equation}
where~$\mu_D$ is the reduced mass of the dark fermions.
The redshift of dark decoupling is determined by~$B_D / \xi$: stronger binding energies and colder initial conditions lead to an earlier and more efficient dark recombination.
In the limit of very large atomic binding energy~$B_D$ and large dark fine-structure constant~$\alpha_D$, dark recombination occurs at very early times, and dark hydrogen behaves as standard CDM particles.
At fixed dark matter abundance, a larger dark proton mass implies a smaller number density.
As long as recombinations keep pace with the expansion rate, the dark ionization fraction follows a Saha-like equilibrium condition,~$x_D^2/(1-x_D) = (\mu_D T_{\gamma,\rm dark}/2\pi)^{3/2} \, e^{-B_D/T_{\gamma,\rm dark}}/n_D$, in which a smaller~$n_D$ sustains a larger equilibrium ionization at a given temperature, since ions encounter recombination partners more rarely.
For the same reason, the recombination rate, itself proportional to~$n_D$, falls below the Hubble rate earlier, freezing the ionized fraction out at a larger value.
A larger dark proton mass therefore leaves a larger residual ionized component.
Thus, the relic ionized fraction is a function of the dark sector microphysics and corresponds to the fraction of millicharged DM in our MAD scenario.

The remaining parameter,~$\epsilon$, controls the strength of the interactions between the two sectors and is the central new ingredient of MAD with respect to ADM.
While in the ADM scenario the dark sector could be entirely sequestered from the visible one, here we are interested precisely in the regime where the kinetic mixing between dark and visible photons allows the free dark fermions to scatter off SM particles, with consequences for cosmological observables.

The primary physical roles of the various parameters are as follows:~$\Delta N_D$ sets the dark radiation abundance; the combination~$B_D/\xi \propto \mu_D \alpha_D^2 / \xi$ controls the redshift of dark recombination; the dark hydrogen mass is the dominant control of the residual ionized fraction left after dark recombination, and it is mainly set by the dark proton mass~$m_{p_D}$, since we take~$m_{p_D} \gg m_{e_D}$; and~$\epsilon$ sets the DM-baryon interaction strength.

\section{DM-SM interactions}
\label{sec:interactions}

The free, ionized fractions of SM fermions $B=\{p,e\}$ and dark fermions $D=\{p_D,e_D\}$ scatter with each other through a $t$-channel process, with a matrix element scaling as~$\epsilon e^2$.
Scatterings of dark ions off the SM photon and loop processes involving dark photon-photon scatterings scale as~$\epsilon^2e^2$ and~$\epsilon^2 e^2 g^2$, respectively, and we neglect these suppressed higher-order interactions.
We also assume that the dark hydrogen is neutral with respect to the visible photon, neglecting residual electromagnetic interactions between the dark bound state and SM particles.

Because the interaction is mediated by a massless photon, the cross section exhibits the familiar Rutherford~$v^{-4}$ enhancement at low velocities, regulated in the plasma by Debye screening~\cite{Dvorkin:2013cea,Nguyen:2021cnb}.
The~$v^{-4}$ scaling results in DM-baryon scattering becoming more important at late times.

The momentum-transfer cross section for elastic scattering between $D$ and $B$ particles is
\begin{equation}
\label{eq:momentum_transfer_cross_section}
    \sigma_{DB}(v) = \frac{2\pi\epsilon^2\alpha_{\rm EM}^2}{\mu_{DB}^2v^4} \log \left\{ \frac{4 \mu_{DB}^2 v^2}{ 4\pi \alpha_{\rm EM} } \left[ \frac{n_p+n_e}{T_b} +\frac{\epsilon^2 (n_{p_D}+n_{e_D})}{T_{b,\rm dark}} \right]^{-1} \right\} \, ,
\end{equation}
where~$v$ is the relative velocity of the colliding particles,~$\mu_{DB}=m_Dm_B/(m_D+m_B)$ is the reduced mass,~$n_{B,D}$ are the background number densities of free charged particles, and~$T_b$,~$T_{b,\rm dark}$ are the baryon and dark baryon temperatures, respectively.
The forward-scattering divergence is regulated by the Debye logarithm, derived in Appendix~\ref{app:Debye}, in the presence of a dark plasma.

The coupling in Eq.~\eqref{eq:momentum_transfer_cross_section} is the same for all four scattering channels---both dark fermions carry millicharge~$\pm\epsilon e$, and both SM fermions charge $\pm e$, and the Debye
logarithm varies only slowly with redshift and channel~\cite{Dvorkin:2019zdi,Nguyen:2021cnb}---so the velocity-independent prefactor~$\sigma_{0,DB} \equiv \sigma_{DB}(v) v^4$ differs between channels only through the reduced mass.
Taking~$\sigma_{0,p_D-p}$ as the
reference, the remaining cross sections follow as
\begin{equation}
    \sigma_{0,DB} = \left( \frac{\mu_{p_D-p}}{\mu_{DB}} \right)^2 \sigma_{0,p_D-p} \, ,
\end{equation}
with lighter pairs having parametrically larger cross sections.

\subsection{Three benchmark scenarios}

The parameter space of the model gives rise to qualitatively distinct cosmological signatures.
Before discussing the details of the interactions, we introduce three benchmark scenarios that we use throughout the paper to organize the discussion:
(i) a \emph{SM-like} scenario, in which the dark sector parameters are of the same order of magnitude as the corresponding SM values;
(ii) an \emph{early-dark-recombination} scenario, in which the dark fermions form bound states well before SM recombination, and dark hydrogen behaves essentially as CDM;
(iii) a \emph{late-dark-recombination} scenario, in which dark recombination occurs not long before SM recombination.
To compare the strength of the millicharge across these benchmarks, we use a dimensionless parameter~$\ratioDMSMtoDMDR$, defined formally in Sec.~\ref{sec:R}, that quantifies the relative importance of DM-SM versus intra-dark-sector interactions at the time of dark recombination; smaller~$\ratioDMSMtoDMDR$ corresponds to a weaker millicharge coupling.

\begin{figure}
    \centering
    \includegraphics[width=0.4\linewidth]{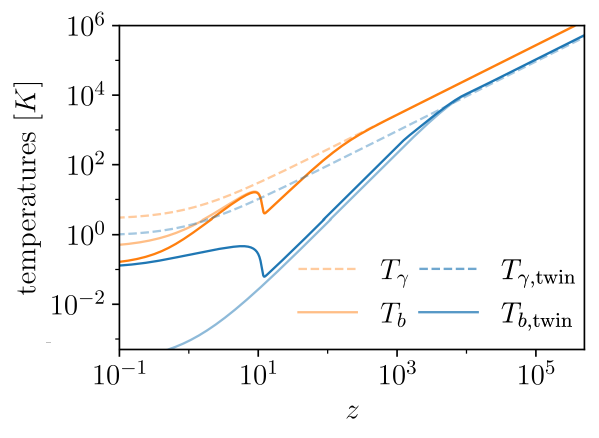}
    \hspace{0.1\linewidth}
    \includegraphics[width=0.4\linewidth]{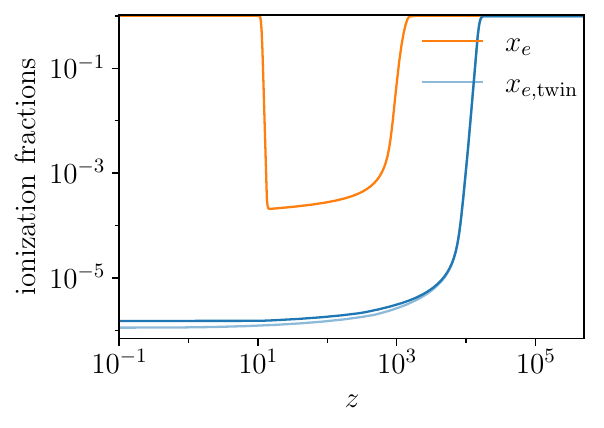}
    \caption{Evolution of temperatures (\emph{left}) and ionization fractions (\emph{right}) as functions of redshift of the visible and dark sectors, with and without the millicharged DM-SM interactions, for an example choice of ADM parameters,~$\{ \Delta N_D = 0.3, \, \alpha_D = 0.01, \, m_{p_D} = 1 \, {\rm GeV}, \, m_{e_D} = 1 \, {\rm MeV} \}$.
    The DM-SM interactions correspond to a millicharge~$\epsilon=10^{-4}$, which for this combination of ADM parameters, translates to large cross section values, chosen here for visualization purposes only, of~$2 \times 10^{-37} \, {\rm cm}^2$ for~$p_D-p$ scatterings, around~$3 \times 10^{-31} \, {\rm cm}^2$ for~$p_D-e$ and~$e_D-e$ scatterings, and~$5 \times 10^{-32} \, {\rm cm}^2$ for~$e_D-p$ scatterings.}
    \label{fig:ex}
\end{figure}

Figure~\ref{fig:ex} illustrates the main thermal effect of the millicharge: it primarily reshapes the \emph{late-time} thermal history of the dark sector, while leaving the early Universe nearly unchanged.
After dark recombination, the dark sector would normally cool adiabatically as~$(1+z)^2$; a sizable millicharge keeps the dark baryons in contact with the (warmer) visible baryons, and the dark sector even heats up following SM reionization, when free SM electrons and protons become abundant.
This behavior is particularly relevant for 21-cm observations at Cosmic Dawn.

\begin{figure}
    \centering
    \includegraphics[width=0.9\linewidth]{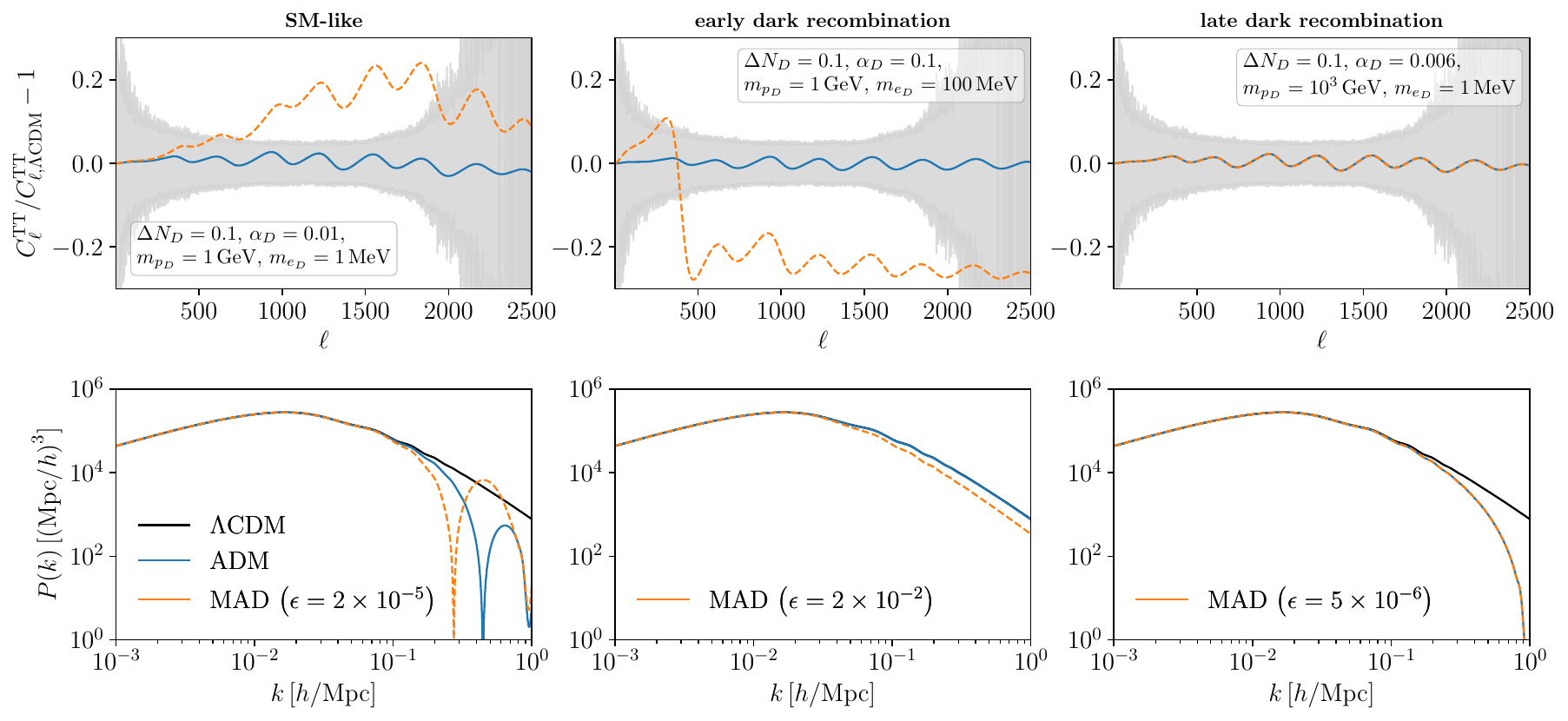}
    \caption{CMB temperature angular power spectrum (\emph{upper row}) and present-day linear matter power spectrum (\emph{bottom row}) for our three benchmark scenarios:
    SM-like (\emph{left}), early dark recombination (\emph{middle}), and late dark recombination (\emph{right}).
    The values of the millicharge are chosen so that, in all three scenarios, they correspond to similar ratios of DM-SM to DM-DR interactions at the moment of dark decoupling ($\mathcal{R} \sim 10^{-4}-10^{-3}$, the parameter~$\mathcal{R}$ is introduced in Sec.~\ref{sec:thermal_history}).
    The largest cross sections (those for~$e_D-e$ scatterings) are of order~$2 \times 10^{-32} \, {\rm cm}^2$ in the SM-like case,~$8 \times 10^{-27} \, {\rm cm}^2$ in the early dark recombination case, and~$10^{-33} \, {\rm cm}^2$ in the late dark recombination case.}
    \label{fig:obs}
\end{figure}

Figure~\ref{fig:obs} previews the corresponding effect on the two main early-Universe observables, the CMB temperature angular power spectrum and the matter power spectrum, comparing each scenario with and without the millicharge.
The qualitative pattern, which we discuss in detail in Sec.~\ref{sec:thermal_history} once the perturbation evolution has been introduced, can already be summarized here: in the SM-like case, even modest interactions modulate the DAOs and shift the CMB peaks enough to rule out the scenario; in the early-dark-recombination case, the millicharged dark fraction behaves analogously to interacting dark matter with baryons, exhibiting the characteristic enhancement of the first CMB peak and suppression of the subsequent ones~\cite{Chen:2002yh,Dvorkin:2013cea,Boddy:2018kfv,Boddy:2018wzy,Xu:2018efh}; in the late-dark-recombination case, the interactions barely modify the pure-ADM signal.
This last case anticipates the existence of regions of parameter space in which MAD is nearly invisible to the CMB and matter power spectrum, yet leaves significant imprints on late-time temperature observables---a possibility we quantify in Sec.~\ref{sec:thermal_history}.

\subsection{Momentum-transfer and heat-transfer rates}

The momentum-transfer and heat-transfer rates are obtained from the momentum-transfer cross section as described in Refs.~\cite{Munoz:2015bca,Boddy:2018wzy}.
The momentum-transfer rate is
\begin{equation}
\label{eq:momentum_transfer_rate}
    R_{DB} = a \frac{\rho_B\sigma_{0,DB}\mathcal{N}_{-4}}{(m_D+m_B) u_{DB}^3} \, ,
\end{equation}
where~$\rho_B = \rho_b (1-Y_{\rm He}) x_e \left( m_B / m_{\rm H} \right)$ is the energy density of the baryonic species participating in the scattering,~$Y_{\rm He}$ is the primordial helium fraction,~$u_{DB}$ is the relative thermal velocity dispersion, and~$\mathcal{N}_{-4} = \frac{1}{3}\sqrt{\frac{2}{\pi}}$.
We model the thermal velocity dispersion as~\cite{Dvorkin:2013cea}
\begin{equation}
\label{eq:thermal_velocity_dispersion}
    u_{DB}^2 = \frac{T_b}{m_B} + \frac{T_{b,\rm dark}}{m_D} + \frac{V_{\rm rms}^2}{3} \, ,
\end{equation}
where the effect of bulk velocity between baryons and dark matter is captured by~$V_{\rm rms}\sim 30\,{\rm km/s}$ at~$z \gtrsim 10^3$ and dilutes with the expansion of the Universe as~$(1+z)$ at later times~\cite{Tseliakhovich:2010bj}.
This mean-velocity prescription coincides with the exact linear-order treatment of DM-baryon scattering for CMB observables~\cite{Ali-Haimoud:2023pbi}.

The corresponding heat-transfer rate---the rate at which scatterings with the free~$B$ bath drive the temperature of species~$D$ toward~$T_b$---entering the temperature evolution equations is
\begin{equation}
    R_{DB}' = \frac{m_D}{m_D+m_B} R_{DB} \, ,
\end{equation}
where the mass ratio reflects the kinematics of elastic scattering: momentum is redistributed at the rate~$R_{DB}$, but the energy exchanged per collision is suppressed for~$m_D \ll m_B$, so a light particle scattering off heavier partners experiences drag from momentum transfer well before it thermalizes.

\section{Thermal histories}
\label{sec:thermal_history}

\subsection{Temperature evolution}

We refer the reader to Refs.~\cite{Cyr-Racine:2012tfp,Bansal:2022qbi} for an in-depth treatment of the thermal history of ADM.
Here, we focus on the evolution equations for the baryon and dark baryon temperatures, which are the most useful tool to keep track of the interplay among the various interaction rates.

Several processes can contribute to maintaining or breaking thermal equilibrium in the dark plasma.
Unlike the SM, where Thomson scattering between photons and electrons dominates, ADM admits a wider set of relevant rates summarized in Table~\ref{tab:rates}.
Bremsstrahlung, photo-recombination, and photo-ionization can dominate over the dark Compton rate when the dark sector is cold or weakly coupled; Rayleigh scattering between dark photons and dark hydrogen becomes relevant when the dark ionization fraction is low and the dark fermion mass ratio is moderate.
On top of these intra-sector processes, MAD adds the heat-exchange rate~$\RDMSMdark'$ and the momentum-transfer rate~$R_{\rm DM-SM}$, which couple the dark and visible sectors through the residual dark ionized fraction.

\begin{table}[t]
    \centering
    \begin{tabular}{|l|l|l|}
        \hline
        Symbol & Process & Role \\
        \hline
        $\Gamma_{\rm C}$ & SM Compton scattering ($\gamma$-$e$) & Couples visible photons and SM baryons \\
        $\Gamma_{\rm C,dark}$ & Dark Compton scattering ($\gamma_D$-$e_D$) & Couples dark photons and dark baryons \\
        $\Pi_{\rm ff}$ & Bremsstrahlung in the dark sector & DM-DR heat exchange when dark plasma is cold or weakly coupled \\
        $\Pi_{\rm p-r}$, $\Pi_{\rm p-i}$ & Dark photo-recombination/ionization & DM-DR heat exchange around dark recombination \\
        $\Pi_{\rm R}$ & Dark Rayleigh scattering & DM-DR heat exchange at low dark ionization fraction \\
        \hline
        $\RDMSMdark'$, $\RDMSM'$ & DM-SM heat-transfer rates & Cross-talk between dark and visible sector temperatures (Eq.~\eqref{eq:total_heat_exchange_rate}) \\
        $R_{\rm DM-SM}$ & DM-SM momentum-transfer rate & Cross-talk in the perturbation equations (Eq.~\eqref{eq:total_momentum_transfer_rate}) \\
        \hline
    \end{tabular}
    \caption{Description of interaction rates. The first block lists the intra-sector processes that keep each plasma in equilibrium; the second block lists the new DM-SM rates introduced by the millicharge.}
    \label{tab:rates}
\end{table}

The temperature evolution equations for the visible and dark sectors then read\footnote{We note that the Bremsstrahlung heating/cooling terms in Refs.~\cite{Cyr-Racine:2012tfp,Bansal:2022qbi} contain a typographical error in the definition given in their appendices; the corresponding implementation in the CLASS ADM code is unaffected.}
\begin{align}
    \frac{dT_b}{dz} &= \frac{1}{1+z} \left[ 2T_b +\frac{\Gamma_{\rm C}}{H}(T_b-T_{\gamma}) +\frac{2 \RDMSM'}{H}(T_b-T_{b,\rm dark}) \right] \, , \\
    \frac{dT_{b,\rm dark}}{dz} &= \frac{1}{1+z} \left[ 2T_{b,\rm dark} +\frac{2 \left(\Pi_{\rm p-r} -\Pi_{\rm p-i} -\Pi_{\rm ff} -\Pi_{\rm R}\right)}{3k_B n_{b,\rm dark} H} +\frac{\Gamma_{\rm C,dark}}{H}(T_{b,\rm dark}-T_{\gamma,\rm dark}) +\frac{2 \RDMSMdark'}{H}(T_{b,\rm dark}-T_b) \right] \, ,
\end{align}
where the baryon number density is~$n_b = n_{\rm H} (1+x_e+f_{\rm He})$, i.e.,~all the particles that share~$T_b$, with~$n_{\rm H}$ the hydrogen nuclei density and~$f_{\rm He} = n_{\rm He}/n_{\rm H}$; similarly, the dark baryon number density is~$n_{b,\rm dark} = n_{{\rm H}_D}(1+x_{e,\rm dark})$, where~$n_{{\rm H}_D}$ counts dark hydrogen nuclei free and bound. 
The first term in each evolution equation describes adiabatic cooling with the expansion of the Universe.
The second term encodes interactions with the respective radiation bath.
The radiation component redshifts as~$(1+z)$, for both SM and dark sectors.\footnote{The DR receives a correction factor that accounts for the possibility of relativistic dark electrons. In practice, in our parameter space, this factor is always unity even at the initial redshift employed by the code.}
The explicit expressions of the DM-DR rates are given in Refs.~\cite{Cyr-Racine:2012tfp,Seager:1999km,Bansal:2022qbi} and reported in Appendix~\ref{app:TCAs} for completeness.
Thermal equilibrium in the dark sector is preserved as long as any of the DM-DR processes has a rate larger than Hubble.
Following Ref.~\cite{Bansal:2022qbi}, we neglect the photo-recombination and photo-ionization terms~$\Pi_{\rm p-r}$,~$\Pi_{\rm p-i}$, which have a similar parametric dependence to the Bremsstrahlung rate; this approximation saves computational cost without significant changes to the thermal history of the dark sector.

Finally, the last term in both equations captures the cross-talk between the two sectors.
The heat exchange rates between DM and SM free ionized particles are
\begin{equation}
\label{eq:total_heat_exchange_rate}
    \RDMSMdark' = \frac{n_b}{n_{b,\rm dark}} \RDMSM' = \frac{1}{a} \sum_{DB} \frac{n_{D,\rm free}}{n_{b,\rm dark}} R_{DB}' \, .
\end{equation}
These expressions differ from the standard DM-baryon-scattering ones, because the sub-species of the DM have different masses; the factors of~$n_{D,\rm free}/n_{b,\rm dark}$ keep track of the number density of species participating in the scattering compared to the whole species sharing the same temperature.
More details on these ratios are given in Appendix~\ref{app:temperature_eqs}.

\subsection{An effective interaction parameter}
\label{sec:R}

Parameterizing the DM-SM interaction in terms of the cross section~$\sigma_{0,p_D-p}$ has limited utility for our purposes, because the physical impact of a given cross section strongly depends on when dark recombination occurs and how efficient it is.
The same cross-section amplitude can correspond to negligible or drastic changes in the final observables, depending on the rest of the parameters.
Moreover, very large interaction rates at early times can spoil the DM-DR tight-coupling approximation and the assumptions underlying dark recombination.

We therefore introduce an effective parameter with a physical interpretation that is more transparent: the relative importance of DM-SM to DM-DR interactions at the time of dark recombination.
We define the ratio of the DM-SM heat-exchange rate to the dark Compton rate, evaluated at (thermal) dark decoupling,
\begin{equation}
    \left. \frac{2 \RDMSMdark'}{\Gamma_{\rm C,dark}} \right|_{z_{\rm dec,dark}} \equiv \ratioDMSMtoDMDR \, .
\end{equation}
This ratio automatically absorbs the dependence on the dark sector microphysics: a given value of~$\ratioDMSMtoDMDR$ corresponds to weakly perturbative DM-SM scatterings during dark recombination, irrespective of the underlying parameters.
We enforce that DM-DR tight coupling holds at early times by restricting our parameter space to~$\ratioDMSMtoDMDR < 10^{-3}$.
We stress that~$\ratioDMSMtoDMDR$ is not designed to map uniquely onto late-time observables; its role is to quantify how perturbative the millicharge coupling is with respect to the internal dark-sector dynamics.
When considering a given value of~$\sigma_{0,DB}$, the boundary of validity of our treatment depends on all the other model parameters, whereas it is a single threshold in terms of~$\ratioDMSMtoDMDR$.
This is what makes~$\ratioDMSMtoDMDR$ the natural sampling variable in our analysis, where our DM-DR tight-coupling requirement becomes a trivial prior bound.

To derive an analytical estimate of the ratio, we assume that DM-DR tight coupling is driven by dark Compton scattering.\footnote{This assumption holds true in the region of parameter space that yields relevant phenomenology for 21-cm observables. We find that when Rayleigh scattering becomes the dominant heat-exchange channel after dark recombination, the resulting dark ionized fraction drops to very low values~$x_{e,\rm dark}(z=0) \lesssim 10^{-6}$~\cite{Cyr-Racine:2012tfp}.}
We find
\begin{equation}
    \Gamma_{\rm C,dark} \simeq \left[1+\left(\frac{m_{e_D}}{m_{p_D}}\right)^3\right] \frac{4 \sigma_{\rm dark} \rho_{0,\gamma,\rm dark}}{3 m_{e_D} c} (1+z)^4 \, ,
\end{equation}
and
\begin{equation}
    2\RDMSMdark' \simeq \sigma_{0,p_D-p} \mu_{p_D-p}^2 \mathcal{N}_{-4} \frac{\rho_{0,b} (1-Y_{\rm He})}{m_{\rm H}} \sum_{DB} \frac{1}{m_D m_B} \left( \frac{T_{0,\gamma}}{m_B} + \frac{T_{0,\gamma,\rm dark}}{m_D} \right)^{-3/2} (1+z)^{3/2} \, ,
\end{equation}
where we used the fact that in our parameter space dark recombination occurs prior to SM recombination, to set~$x_e \sim 1$ and~$T_b \sim T_{\gamma}$.

The ratio~$\ratioDMSMtoDMDR$ is evaluated at the estimated redshift of thermal decoupling~$z_{\rm dec,dark}^{\rm (estimate)}$, which is obtained in the code by bisection, following the implementation of \CLASSADM, as the redshift at which the dark Compton heating rate, evaluated with the Saha ionization fraction, drops below~$100$ times the Hubble rate (the factor~$100$ marks the onset of thermal decoupling).
By construction, this estimate of~$\ratioDMSMtoDMDR$ holds down to the redshift of dark recombination and immediately breaks down at lower redshifts.

\begin{figure}
    \centering
    \includegraphics[width=0.9\linewidth]{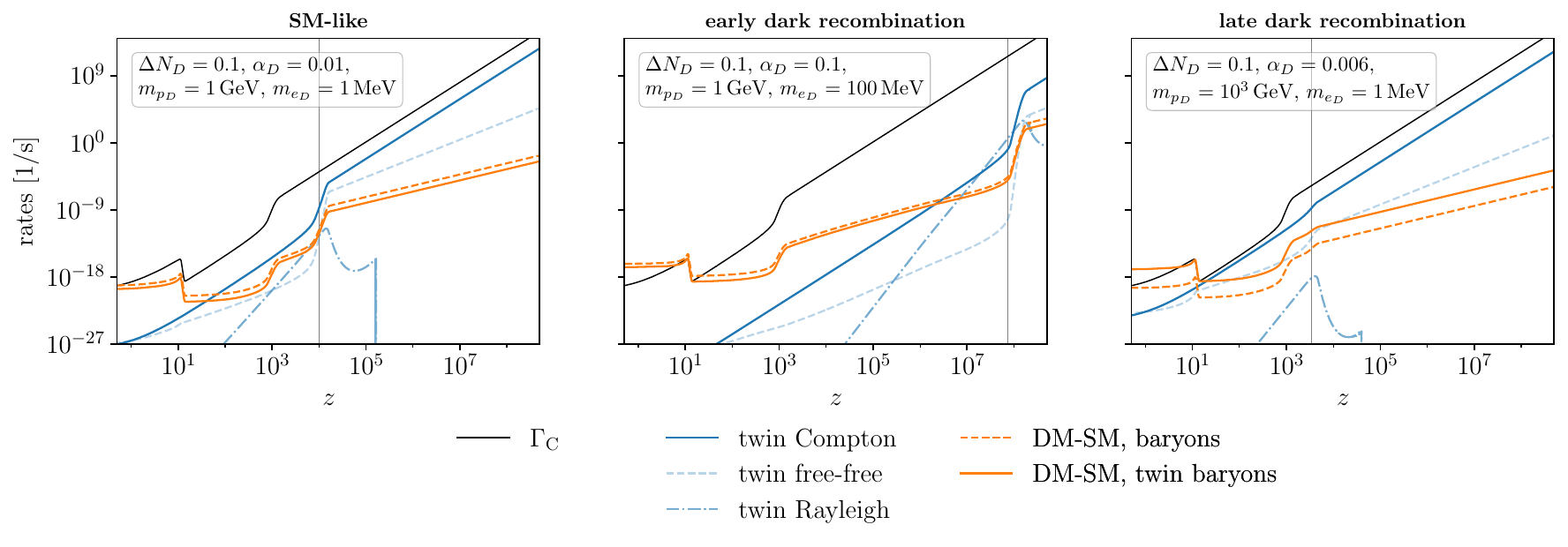}
    \caption{Redshift evolution of the rates of Table~\ref{tab:rates} for the three benchmarks of Figure~\ref{fig:thermal_histories}. The values of the cross sections in the three scenarios yield similar values of~$\ratioDMSMtoDMDR \sim 10^{-3}$ in all three cases (compare the orange solid line to the solid blue line). The redshift of dark decoupling is marked by the gray vertical line.}
    \label{fig:rates}
\end{figure}

The benchmarks introduced in Sec.~\ref{sec:interactions}---SM-like, early-dark-recombination, late-dark-recombination---allow us to illustrate the interplay among the various rates.
Figure~\ref{fig:rates} shows their redshift evolution.
The DM-SM interaction cross sections in the three scenarios have been chosen so that~$\ratioDMSMtoDMDR \sim 10^{-3}$.
Before dark decoupling (gray vertical line), the dark sector is dominated by dark Compton scattering and the DM-DR tight-coupling approximation holds.
After dark decoupling, Bremsstrahlung, Coulomb scattering, and the DM-SM rates compete differently across benchmarks: in the early-dark-recombination case the DM-SM rate is comparatively much larger at low redshifts, while in the late-dark-recombination case the dark dynamics remain dominated by internal DM-DR processes well past SM recombination.
A similar value of~$\ratioDMSMtoDMDR$ thus encodes physically distinct regimes: this is by construction, since~$\ratioDMSMtoDMDR$ measures the perturbativity of the DM-SM coupling during dark recombination, not the size of its late-time imprints.
The imprints themselves are governed by a small set of parameter combinations that translate the model parameters into the quantities cosmology responds to: $\Delta N_D$ sets the dark radiation density and its effects on the expansion history; $B_D/\xi$ sets the redshift of dark recombination, which determines which scales carry dark acoustic oscillations and whether the dark sector is already CDM-like by CMB times; the relic ionized fraction~$x_D$, fixed by~$\xi$, $\alpha_D$, and the dark fermion masses (Sec.~\ref{sec:model}), plays the role of the millicharged fraction; and~$\epsilon$, together with~$x_D$ and the dark fermion masses, sets the strength of the late-time DM-SM energy and momentum exchange.

\begin{figure}
    \centering
    \includegraphics[width=0.9\linewidth]{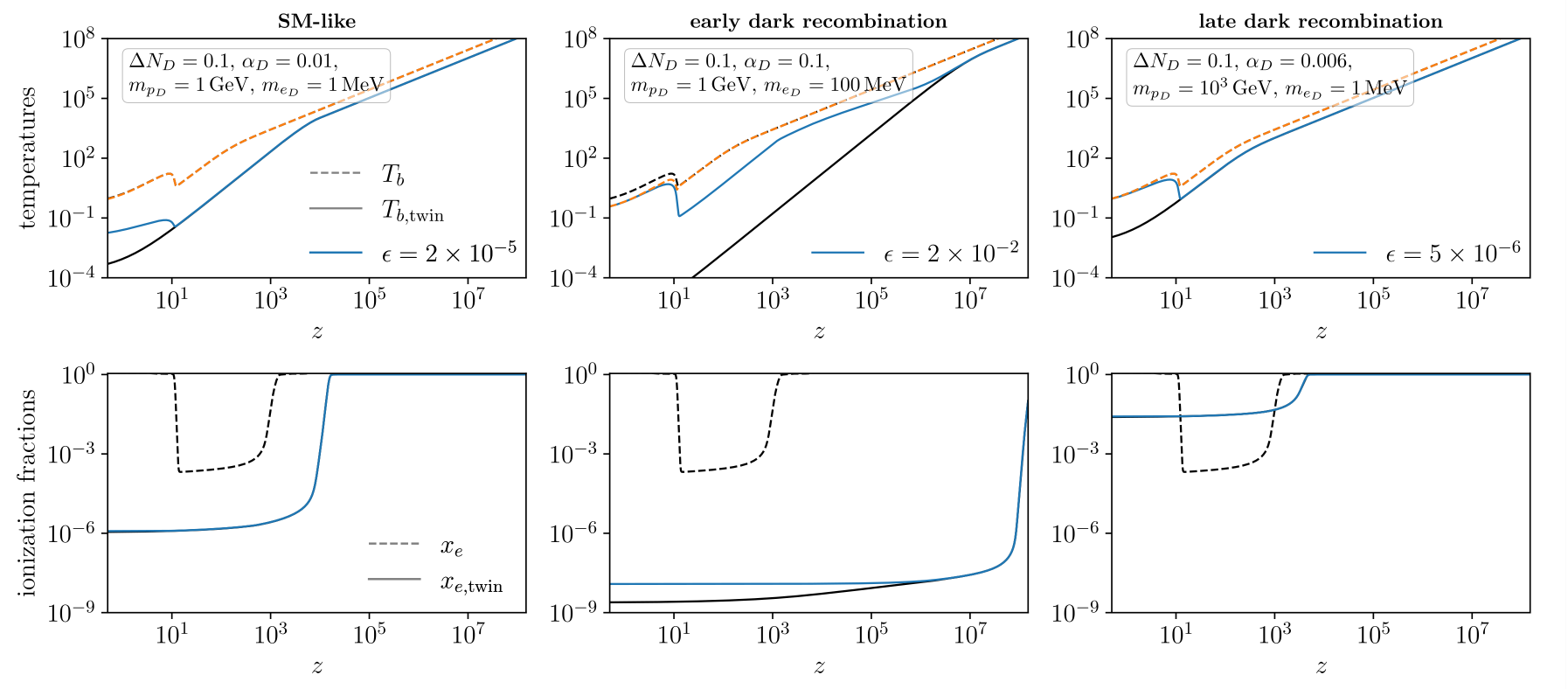}
    \caption{Temperatures (\emph{top}) and ionization fractions (\emph{bottom}) for the three benchmarks of Figure~\ref{fig:obs}.
    Dashed lines: baryon temperature; solid lines: dark baryon temperature.
    Black: ADM with no millicharge.
    Orange and blue: MAD with the same millicharge values as in Figure~\ref{fig:obs} (orange for the baryons, blue for the dark baryons).
    The MAD realization in the two leftmost columns corresponds to interaction cross sections already ruled out by \textit{Planck}; the right column is the regime of interest for late-time probes.}
    \label{fig:thermal_histories}
\end{figure}

Figure~\ref{fig:thermal_histories} shows the corresponding temperatures and ionization fractions for ADM (black) and for MAD, with the same millicharge values as Figure~\ref{fig:obs}, corresponding to~$\ratioDMSMtoDMDR \sim 10^{-3}$.
Elastic Coulomb-like scatterings between the residual free dark fermions and the free SM electrons and protons transfer heat from the warmer baryons to the colder dark sector: after dark recombination the dark baryons would otherwise cool adiabatically, and these collisions drag~$T_{b,\rm dark}$ back up toward~$T_b$.
From the baryon side, the collisions act as a heat sink: at high coupling, the dark sector cools the baryons appreciably.
Millicharge values this large are already ruled out by \textit{Planck} in the SM-like and early-dark-recombination benchmarks; the late-dark-recombination benchmark, by contrast, allows sizable shifts of the dark baryon temperature while leaving the CMB intact.
This is the regime targeted by 21-cm observations.
The corresponding cross sections~$\sigma_{0,DB}$ are reported in Table~\ref{tab:corresponding_cross}.
Interactions between the lightest particles have the largest cross sections, but the interaction rates appear weighted by number densities in the temperature equations and by energy densities in the perturbation equations (see Appendices~\ref{app:temperature_eqs}–\ref{app:perturbations_eqs}), so the channel hierarchy in the observables is not the same as in the cross sections themselves.

\begin{table}[]
    \centering
    \begin{tabular}{l|c|c|c}
         & \multicolumn{1}{c|}{\hspace{1.3cm}SM-like\hspace{1.1cm}\,} & \multicolumn{1}{c|}{\hspace{0.1cm}early dark recombination\hspace{0.05cm}\,} & \multicolumn{1}{c}{\hspace{0.1cm}late dark recombination\hspace{0.05cm}\,} \\
        \hline
        $\epsilon$ & $2 \times 10^{-5}$ & $2 \times 10^{-2}$ & $5 \times 10^{-6}$ \\
        \hline
        $\sigma_{0,p_D-p} \, [{\rm cm}^2]$ & $9.50 \times 10^{-39}$ & $9.30 \times 10^{-33}$ & $1.60 \times 10^{-40}$ \\
        $\sigma_{0,p_D-e}$ & $8.52 \times 10^{-33}$ & $8.35 \times 10^{-27}$ & $5.39 \times 10^{-34}$ \\
        $\sigma_{0,e_D-p}$ & $2.23 \times 10^{-33}$ & $2.67 \times 10^{-31}$ & $1.41 \times 10^{-34}$ \\
        $\sigma_{0,e_D-e}$ & $1.94 \times 10^{-32}$ & $8.43 \times 10^{-27}$ & $1.23 \times 10^{-33}$
    \end{tabular}
    \caption{Millicharges and cross sections corresponding to the example thermal histories shown in Figures~\ref{fig:obs},~\ref{fig:rates}, and~\ref{fig:thermal_histories}.}
    \label{tab:corresponding_cross}
\end{table}

\subsection{Cosmological perturbations}

The evolution of the SM baryon and dark baryon cosmological perturbations for density and velocity divergence in conformal Newtonian gauge, in the customary formalism of~\cite{Ma:1995ey}, is given by
\begin{align}
    & \Dot{\delta}_{b,\rm dark} + \theta_{b,\rm dark} -3 \Dot{\phi} = 0 \, , \\
    & \Dot{\theta}_{b,\rm dark} +\frac{\Dot{a}}{a} \theta_{b,\rm dark} -c_D^2 k^2 \delta_{b,\rm dark} -k^2 \psi \nonumber \\
    & \hspace{1.2cm} = -\frac{4\rho_{\gamma,\rm dark}}{3\rho_{b,\rm dark}} \frac{1}{\tau_D} \left( \theta_{b,\rm dark} - \theta_{\gamma,\rm dark} \right) -R_{\rm DM-SM} \left( \theta_{b,\rm dark} - \theta_b \right) \, , \\
    & \Dot{\delta}_b + \theta_b -3\Dot{\phi} = 0 \, , \\
    & \Dot{\theta}_b +\frac{\Dot{a}}{a} \theta_b -c_s^2 k^2 \delta_b -k^2 \psi = -\frac{4\rho_{\gamma}}{3\rho_b} \frac{1}{\tau} \left( \theta_b - \theta_{\gamma} \right) +\frac{\rho_{b,\rm dark}}{\rho_b} R_{\rm DM-SM} \left( \theta_{b,\rm dark} - \theta_b \right) \, ,
\end{align}
where dots denote derivatives with respect to conformal time,~$\phi$ and~$\psi$ are the metric perturbations, and~$c_s \, (c_D)$ and~$\tau \, (\tau_D)$ denote the (dark) sound speed and the opacity of the (dark) plasma~\cite{Bansal:2022qbi}.
The DM-SM momentum transfer rate is
\begin{equation}
\label{eq:total_momentum_transfer_rate}
    R_{\rm DM-SM} = \frac{\rho_{e_D,\rm free}}{\rho_{b,\rm dark}} \left(R_{e_D-e}+R_{e_D-p}\right) + \frac{\rho_{p_D,\rm free}}{\rho_{b,\rm dark}} \left(R_{p_D-p}+R_{p_D-e}\right) \, ,
\end{equation}
where the individual momentum transfer rates~$R_{DB}$ are those given in Eq.~\eqref{eq:momentum_transfer_rate}.
The weights by energy density ratios are derived in Appendix~\ref{app:perturbations_eqs}, analogously to how the familiar~$\rho_{\gamma}/\rho_b$ factor appears in the photon-baryon perturbation equations.
They arise from summing the individual moments of the Boltzmann equations for free dark electrons, free dark protons, and dark hydrogen atoms, requiring that their energy densities sum to~$\rho_{b,\rm dark}$ and that they share a common velocity.

\smallskip
\noindent\emph{The temperature evolution is weighted by number-density ratios; the perturbation evolution, by energy-density ratios. Temperatures track which species participates in the heat exchange; perturbations track which species carries momentum.}
\smallskip

\begin{figure}
    \centering
    \includegraphics[width=0.9\linewidth]{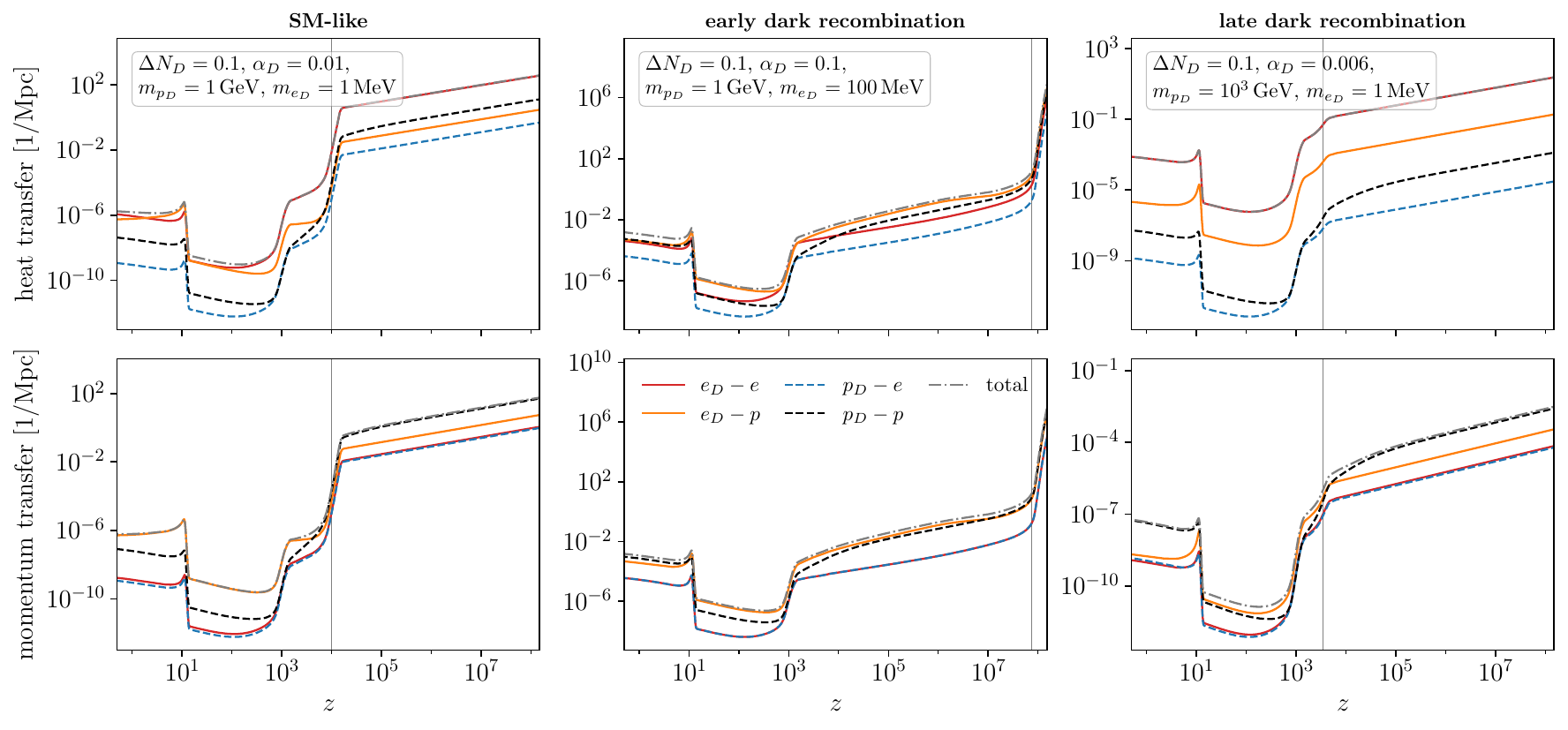}
    \caption{Decomposition of the heat exchange rate~$a \RDMSMdark'$ (\emph{top}, Eq.~\eqref{eq:total_heat_exchange_rate}) and the momentum exchange rate~$R_{\rm DM-SM}$ (\emph{bottom}, Eq.~\eqref{eq:total_momentum_transfer_rate}) into the individual DM-SM scattering channels, for the three benchmarks of Figure~\ref{fig:thermal_histories}.
    The redshift of dark decoupling is marked by the gray vertical line.
    The features in the rates correspond to dark recombination, SM recombination, and SM reionization, through the evolution of the free ionized fractions.}
    \label{fig:heat_mom_transfer}
\end{figure}

This decoupling has a concrete consequence: the same set of cross sections impacts temperatures and perturbations through different combinations of processes.
Figure~\ref{fig:heat_mom_transfer} illustrates this by decomposing the total heat and momentum exchange rates into the four scattering channels.
Heat exchange is generically dominated by the lighter dark electrons (because they carry most of the thermal energy per particle), whereas momentum exchange is dominated by the heavier dark protons.
This is one reason why scenarios that appear similar in the temperature evolution can still leave distinguishable signatures in the matter power spectrum, and vice versa.

With the temperature and perturbation evolution now in hand, we can return to Figure~\ref{fig:obs}, previewed in Sec.~\ref{sec:interactions}, and describe the three columns in more detail.
In the \emph{SM-like} column, the dark fermions reach dark recombination relatively late, so a sizable population of free dark ions persists into epochs when DM-SM momentum transfer can still influence cosmological perturbations; even modest interactions then modulate the DAOs and shift the CMB peaks enough to be excluded.
In the \emph{early-dark-recombination} column, dark recombination has long completed by the time of SM recombination, and the residual ionized fraction acts like a subpercent component of interacting dark matter with baryons: the CMB shows the characteristic enhancement of the first acoustic peak and suppression of the higher-$\ell$ oscillations, while the matter power spectrum is damped on small scales.
In the \emph{late-dark-recombination} column, internal dark-sector processes still dominate the dark-baryon dynamics at the relevant redshifts, so the millicharge interactions leave only a faint imprint on the perturbations; here the action lives in the late-time thermal history rather than in the early-Universe observables.

After dark recombination, the ADM sector can be modeled in an effective way as a sector of interacting dark matter and dark radiation, within the ETHOS formalism~\cite{Cyr-Racine:2015ihg}, where the opacity of the dark sector is set by the Thomson, Rayleigh, and photo-ionization scattering rates.
The mapping implemented in the code was developed by Ref.~\cite{Bansal:2022qbi} and is spelled out in Table 1 therein.

\section{Analysis}
\label{sec:analysis}

We now explore the parameter space of the model and compare its predictions to observations of the CMB and large-scale structure.
We compare MAD to the cases of ADM with no millicharge, of $\Lambda$CDM, and of $\Lambda$CDM$+N_{\rm eff}$ in order to isolate the effect of the dark radiation alone.
We perform a Markov Chain Monte Carlo analysis using the code MontePython v3.6.1~\cite{Brinckmann:2018cvx}.
We compute the CMB and matter power spectra with \CLASSMAD and sample the posterior distribution of the MAD parameters, given the following cosmological datasets:
\begin{itemize}
    \item CMB data from \textit{Planck} 2018 for temperature, polarization, and lensing~\cite{Planck:2019nip} (specifically, high-$\ell$ TTTEEE lite, low-$\ell$ EE, low-$\ell$ TT, and lensing),
    \item ACT DR6 lensing data~\cite{ACT:2023dou,ACT:2023kun},
    \item baryon acoustic oscillation (BAO) data from BOSS Data Release 12~\cite{BOSS:2016wmc}, SDSS Data Release 7 MGS~\cite{Ross:2014qpa}, 6dFGS~\cite{Beutler:2011hx}, and DESI~\cite{DESI:2024mwx}, and
    \item Pantheon supernovae~\cite{Brout:2022vxf}.
\end{itemize}
We use the likelihood for ACT and DESI BAO adapted for MontePython by Refs.~\cite{Herold:2024nvk,Herold:2025hkb}.\footnote{The likelihood is available at \href{https://github.com/cosmodesi/desilike}{\lstinline{https://github.com/cosmodesi/desilike}}.}

On top of the $\Lambda$CDM set of cosmological parameters,
\begin{equation}
    \left\{ \omega_{\rm b}, \, \omega_{\rm cdm}, \, h, \, \ln 10^{10} A_s, \, n_s \right\} \, ,
\end{equation}
we vary the five MAD parameters,
\begin{equation}
    \left\{ \Delta N_D, \, \alpha_D, \, m_{p_D}, \, m_{e_D}, \, {\rm log}_{10}\ratioDMSMtoDMDR \right\} \, ,
\end{equation}
and in addition, all the appropriate nuisance parameters for the experiments under consideration.

Following a similar approach to Ref.~\cite{Bansal:2022qbi}, we use uniform priors on~$\Delta N_D \in [10^{-4}, 1]$ and on~$\alpha_D \in [0.006, 0.1]$, and uniform priors on the logarithm of the dark fermion masses~${\rm log}_{10} \left( m_{p_D} / {\rm GeV} \right) \in [0, 3]$ and~${\rm log}_{10} \left( m_{e_D} / {\rm MeV} \right) \in [0, \, 2]$.
With these choices of priors, the redshift of dark recombination is allowed to vary between roughly~$10^4-10^8$.
We set a uniform prior in~${\rm log}_{10} \ratioDMSMtoDMDR \in [-25, -3]$, where the upper bound carves out the region in parameter space where dark recombination is not hindered by DM-SM interactions.

Two features of this prior choice should be flagged upfront, as they shape the interpretation of the posterior.
First, a log-uniform prior on~$\ratioDMSMtoDMDR$ implies a non-uniform prior on the underlying cross section and excludes the exact zero-interaction limit by construction; we adopt it because at very small cross sections the likelihood is essentially flat and a uniform prior would dilute information about the cosmologically interesting regime (e.g.,~see Refs.~\cite{Straight:2026wts, Smith:2020rxx}).
Second, as a consequence, the posterior at sufficiently small cross sections is prior dominated: any apparent ``preference'' for very small~$\ratioDMSMtoDMDR$ should be read as the absence of an upper bound from data, not as a positive detection.
We have chosen the lower bound on~$\ratioDMSMtoDMDR$ small enough that this effect is confined to the lowest-cross-section region, and we have verified that the results at larger cross sections are insensitive to the lower cutoff.

\section{Discussion}
\label{sec:discussion}

As often found in the ADM literature, most two-parameter combinations are nearly unconstrained when marginalizing over the others, except for~$\Delta N_D$~\cite{Bansal:2022qbi}.
This is a mere consequence of the fact that $\Lambda$CDM provides a good fit to the data, and the marginalization process folds into the projection regions of parameter space where the model behaves similarly to $\Lambda$CDM, or to $\Lambda{\rm CDM}+N_{\rm eff}$.
However, we recover the strong bound~$\Delta N_D \lesssim 0.3$, which is required by \textit{Planck} data to accommodate the presence of dark radiation.

The details of dark recombination, i.e.,~the dark recombination redshift and the dark ionization fraction, change by~$< \mathcal{O}(10\%)$ between ADM and MAD.
This is expected, since we work within the requirement that DM-SM interactions should not spoil the process of dark recombination, and it is a check of our procedure.

\subsection{$H_0$ and $S_8$ tensions}

\begin{figure}
    \centering
    \includegraphics[width=0.5\linewidth]{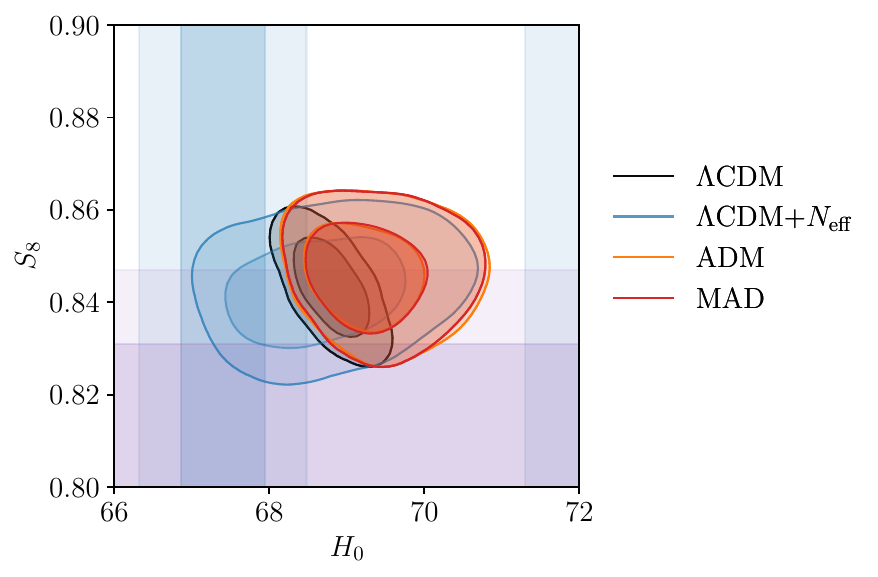}
    \caption{Two-dimensional projection of the posterior in the~$H_0$-$S_8$ plane, after marginalizing over all other parameters; $S_8 = \sigma_8 \sqrt{\Omega_m / 0.3}$.
    Colored bands: $H_0 = 67.36 \pm 0.54$ from \textit{Planck}~\cite{Planck:2018vyg}, $H_0 = 73.5 \pm 1.1$ from Pantheon Plus + SH0ES~\cite{Brout:2022vxf}, $S_8 = 0.815^{+0.016}_{-0.021}$ from KiDS~\cite{Wright:2025xka}.
    Likelihood: \textit{Planck} + ACT lensing + BAO + Pantheon Plus.}
    \label{fig:H0}
\end{figure}

Like ADM, MAD inherits a dark radiation component that can alleviate the Hubble tension~\cite{Garny:2026ish,Bansal:2022qbi,Buen-Abad:2024tlb}, while the DM-DR tight coupling at early times ensures a good fit to the high-$\ell$ CMB.
Small-scale modifications to the matter power spectrum from DAOs additionally make the model a candidate to address the~$S_8$ tension.
Figure~\ref{fig:H0} compares the marginalized posteriors of $\Lambda$CDM, $\Lambda$CDM+$N_{\rm eff}$, ADM, and MAD in the~$H_0$-$S_8$ plane.
MAD and ADM yield nearly identical contours, and both overlap with $\Lambda$CDM+$N_{\rm eff}$.
The ADM and MAD posteriors do not reach low values of $H_0$, because we keep~$N_{\rm eff}=3.044$ fixed and vary~$\Delta N_D$ on top of it, whereas $\Lambda$CDM+$N_{\rm eff}$ allows lower~$N_{\rm eff}$ values that increase the sound horizon before recombination.
The clear conclusion, in agreement with Ref.~\cite{Bansal:2022qbi}, is that the dominant effect of MAD on the~$H_0-S_8$ plane is dark radiation: the millicharged interactions neither significantly improve nor spoil the ability of the model to address these tensions.

\subsection{Cosmological constraints on the millicharge}
\label{subsec:discussion}

\begin{figure}
    \centering
    \includegraphics[width=0.9\linewidth]{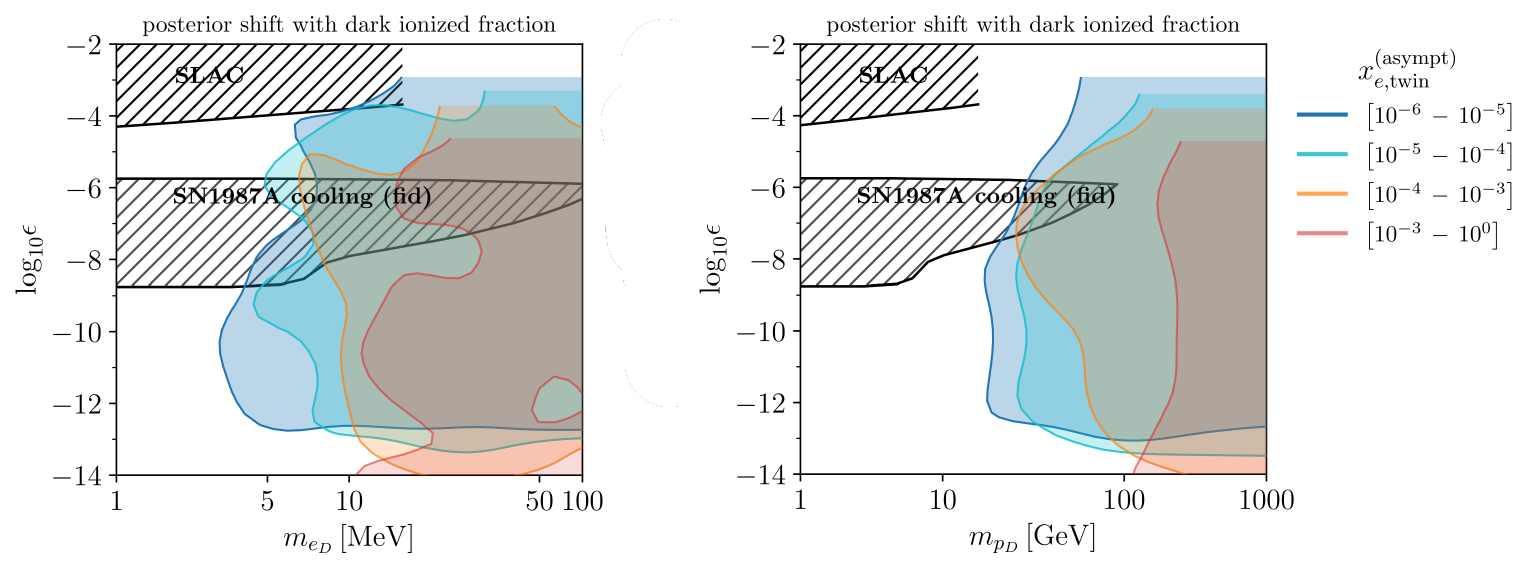}
    \caption{Marginalized posterior contours (68\% C.L.) in the mass-millicharge plane, sliced into bins of the present-day dark ionization fraction~$x_{e,\rm dark}(z=0)$ for the dark electron (\emph{left}) and the dark proton (\emph{right}).
    Hashed regions: SLAC~\cite{Prinz:1998ua} and SN 1987A cooling~\cite{Chang:2018rso} exclusions, independent of the cosmological abundance of millicharged particles.}
    \label{fig:epsilon_posterior}
\end{figure}

The central result of this work is summarized in Figure~\ref{fig:epsilon_posterior}, which illustrates viable MAD realizations under cosmological constraints.
The figure shows the marginalized posterior in the plane spanned by the dark fermion mass and the millicharge~$\epsilon$, sliced into bins of the present-day dark ionization fraction~$x_{e,\rm dark}(z=0)$.
The dark ionization fraction plays the role of the millicharged dark matter fraction and is the natural output of dark recombination.
Unlike in minimal millicharged DM models, the interacting fraction in MAD is not freely chosen but dynamically determined by the dark recombination history.
We focus on millicharged fractions larger than~$10^{-6}$, which is the regime that can affect the 21-cm global signal~\cite{Kovetz:2018zan}.
Our analysis is sensitive to millicharges as low as~$\mathcal{O}(10^{-14})$.

Three external bounds are relevant for comparison.
Lab-based searches for millicharged dark particles constrain~$\epsilon \lesssim 10^{-3}-10^{-1}$ in the mass range~$10 \, {\rm MeV} - 100 \, {\rm GeV}$~\cite{Prinz:1998ua,Alcott:2025rxn}.
Supernova 1987A cooling excludes~$10^{-9} \lesssim \epsilon \lesssim 10^{-6}$ for dark particle masses below~$200 \, {\rm MeV}$~\cite{Chang:2018rso}.
The BBN bound~$\epsilon \lesssim 10^{-9}$~\cite{Davidson:2000hf} only applies to dark particles lighter than the dark constituents considered here.

The posterior contours in Figure~\ref{fig:epsilon_posterior} land in regions complementary to these laboratory and astrophysical exclusions, including the open windows~$10^{-6}\lesssim\epsilon\lesssim10^{-4}$ and~$\epsilon\lesssim 10^{-8}$.
The first of these windows is also the parameter range invoked in the original EDGES-style explanations of baryon cooling at Cosmic Dawn~\cite{Munoz:2015bca,Barkana:2018lgd,Fialkov:2018xre,Munoz:2018pzp,Berlin:2018sjs,Barkana:2018qrx,Kovetz:2018zan,Liu:2019knx}, with relic millicharged fractions of order percent or below: the MAD scenario therefore populates exactly the parameter region that originally motivated millicharged DM as a 21-cm cooling mechanism, but without the need to set the fraction by hand.
A clear physical trend emerges: light dark electron masses correspond to small relic ionized fractions,~$x_{e,\rm dark}(z=0) < 10^{-6}$ which are not shown in the plot, while heavy masses correspond to larger ones.
Around~$m_{e_D} \sim 10 \, {\rm MeV}$ the contours span the broadest range of $x_{e,\rm dark}(z=0)$.
The analogous slicing for the dark proton mass (right panel) shows a similar monotonic correlation: heavier dark protons produce smaller dark hydrogen number densities and thus smaller relic ionized fractions.
Realizations with millicharged fraction~$10^{-6} < x_{e,\rm dark}(z=0) < 10^{-5}$ represent a~$20\%$ of the total sample under the adopted priors,~$10^{-5} < x_{e,\rm dark}(z=0) < 10^{-4}$ represent a~$12\%$,~$10^{-4} < x_{e,\rm dark}(z=0) < 10^{-3}$ a~$6\%$, while cases~$x_{e,\rm dark}(z=0) > 10^{-3}$ only occur in a~$3\%$ of the cases.
Millicharged fractions lower than~$10^{-6}$ are not shown in the plot.

\begin{figure}
    \centering
    \includegraphics[width=0.49\linewidth]{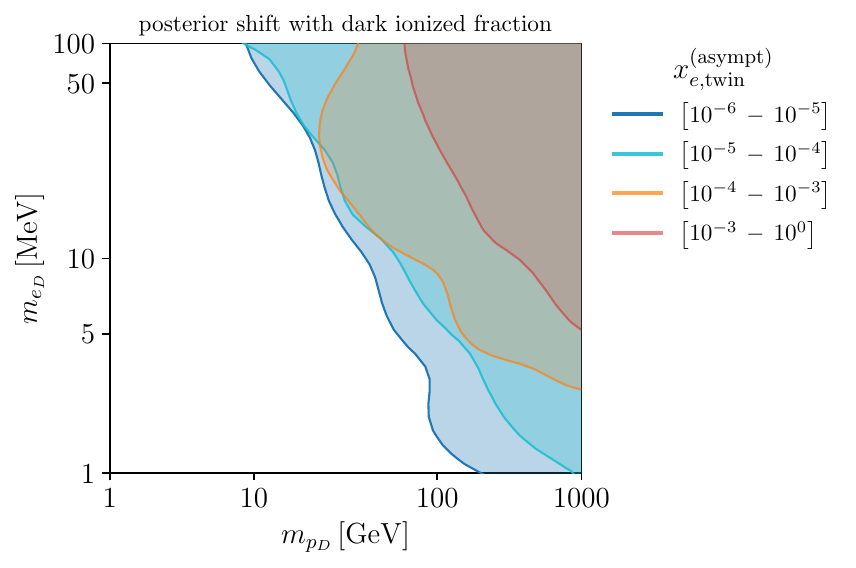}
    \caption{Marginalized posterior slices for~$m_{p_D}$ versus~$m_{e_D}$, with~$x_{e,\rm dark}(z=0)$ bin shown by color, for the same dark ionization fraction bins as Figure~\ref{fig:epsilon_posterior}.}
    \label{fig:posterior_masses}
\end{figure}

To visualize the effect of the dark electron and dark proton masses together, Figure~\ref{fig:posterior_masses} shows the joint distribution in the~$(m_{p_D},m_{e_D})$ plane, color-coded by~$x_{e,\rm dark}(z=0)$ bin.
The final relic density is proportional to the product of the dark hydrogen mass and binding energy,~$x_{e,\rm dark} \propto m_{{\rm H}_D} B_D$, derived from Saha freeze-out of dark recombination~\cite{Cyr-Racine:2012tfp}.
Therefore both the dark constituents masses contribute to the relic fraction: the dark proton mass directly controls the number density of dark hydrogen, while their binding energy is determined by the much lighter dark electron mass.
Indeed, while a heavy dark proton alone is enough to obtain~$x_{e,\rm dark} \gtrsim 10^{-6}$, the largest relic fractions can only be achieved if also the dark electron mass is large enough,~$m_{e_D} \gtrsim 10 \, {\rm MeV}$.

Taken together, Figures~\ref{fig:epsilon_posterior}-\ref{fig:posterior_masses} make the main point of the paper concrete: MAD populates precisely the low-fraction millicharged parameter space that remains cosmologically viable, with the relic ionized fraction mainly controlled by the mass of the heaviest dark constituent.

\section{Conclusions}
\label{sec:conclusions}

For millicharged dark matter to have an interesting effect on 21-cm observations, only a subpercent fraction of the dark matter can be endowed with a fractional electric charge, a regime that minimal millicharge models can occupy by introducing an ad hoc interacting fraction.
Millicharged atomic dark matter (MAD) is a viable model that achieves low fractions of millicharged dark particles (with millicharges that arise from kinetic mixing between the dark and visible photons) as an outcome of the thermal history of a rich dark sector: the small interacting fraction is simply the relic ionization left after dark recombination.
The model belongs to the class of high-energy scenarios with one or more dark sectors related to the SM by a discrete symmetry, such as Twin Higgs models, which are widely studied as solutions to the Hierarchy Problem.
At the same time, atomic dark matter (ADM)-like dark sectors offer routes to alleviate current cosmological tensions, including the Hubble and~$S_8$ tensions, through their dark radiation content and the dark acoustic oscillations imprinted on the matter power spectrum.
MAD therefore sits at the intersection of some of the most pressing open questions in high-energy particle physics and in cosmology, and serves as a useful, physically motivated arena to study them together.

We modified the CLASS code for ADM~\cite{Bansal:2022qbi} to incorporate the millicharged interaction, and we presented the first cosmological constraints on the MAD model from CMB, large-scale structure, BAO, and supernovae data.
Three main findings emerge from the analysis:
\begin{enumerate}
    \item We recover the strong bound~$\Delta N_D \lesssim 0.3$ on the dark radiation contribution; the remaining two-parameter projections of the posterior are largely unconstrained, reflecting both the goodness of fit of $\Lambda$CDM and the flexibility of MAD in approaching it.
    \item The viable mass-millicharge region spans the open windows~$10^{-6} \lesssim \epsilon \lesssim 10^{-4}$ and~$\epsilon \lesssim 10^{-8}$ left by SLAC and SN 1987A, with the relic ionized fraction controlled primarily by the dark proton mass. A broad-compatibility band around~$m_{e_D} \sim 10 \, {\rm MeV}$ accommodates the full range of relic fractions consistent with the data, and overlaps with the parameter window invoked in EDGES-style explanations of baryon cooling at Cosmic Dawn~\cite{Barkana:2018lgd,Munoz:2018pzp,Creque-Sarbinowski:2019mcm}.
    \item On the~$H_0$-$S_8$ plane, MAD performs analogously to ADM, with the millicharge contributing only marginally: the dominant driver in this plane is the dark radiation component.
\end{enumerate}

The cosmologically least-constrained regions of the MAD parameter space are precisely those that leave the largest imprints on the late-time baryon and dark baryon temperatures, which can be probed by upcoming 21-cm line-intensity mapping surveys~\cite{Kovetz:2018zan}.
Combined with the conceptual simplicity of the underlying mechanism---residual dark ionization with a millicharge induced by kinetic mixing---this makes MAD a well-motivated target for the coming generation of cosmological probes.

\begin{acknowledgments}
The authors thank Subhajit Ghosh for useful discussions, and Jared Barron for comments on the \CLASSADM code.
EV and EDK thank the Weinberg Institute at the University of Texas at Austin for the warm hospitality during student and sabbatical visits, respectively.
We acknowledge the Texas Advanced Computing Center (TACC) at The University of Texas at Austin for providing high-performance computing resources.
EV is supported by an Azrieli International Postdoctoral Fellowship.
KB acknowledges support from the National Science Foundation (NSF) under Grant No. PHY-2413016.
EDK acknowledges support from the U.S.--Israel Binational Science Foundation (NSF-BSF grant 2022743 and BSF grant 2024193) and the Israel National Science Foundation (ISF grant 3135/25), as well as from the joint Israel--China program (ISF--NSFC grant 3156/23).
\end{acknowledgments}

\appendix
\import{}{appendix.tex}

\bibliography{bibliography}

\end{document}

%% file: appendix.tex
\section{Cross section, Debye mass and Debye screening}
\label{app:Debye}

The differential cross section for the~$t$-channel scattering process~$D+B \xrightarrow{} D+B$, with~$D=\{p_D,e_D\}$ and~$B=\{p,e\}$, is
\begin{equation}
\begin{split}
    \left(\frac{d\sigma_{DB}}{d\Omega}\right)_{\rm CM} &= \frac{(\epsilon e^2 / 4\pi)^2}{4E_{\rm CM}^2 |\vec{p}|^4 (1-\cos\theta)^2} \left\{ \left[E_{\rm CM}^2-(m_D^2+m_B^2)\right]^2 -2E_{\rm CM}^2 |\vec{p}|^2 (1-\cos\theta) +2|\vec{p}|^4 (1-\cos\theta)^2 \right\} \\
    &\simeq \frac{(\epsilon e^2 / 4\pi)^2 (m_D+m_B)^2}{4 m_D^4 m_B^4 v_{DB}^4 (1-\cos\theta)^2} \left\{ 4 m_D^2 m_B^2 -2 m_D^2 m_B^2 v_{DB}^2 (1-\cos\theta) +2\mu^4 v_{DB}^4 (1-\cos\theta)^2 \right\} \, ,
\end{split}
\end{equation}
where the center-of-mass energy is~$E_{\rm CM} = E_D+E_B \simeq (m_D+m_B)$, the center-of-mass momentum is~$\vec{p} \simeq \mu\vec{v}_{DB}$, and the momentum transfer is~$|\vec{q}|^2 = 2 |\vec{p}|^2 (1-\cos\theta) = 2 \mu^2 v_{DB}^2 (1-\cos\theta)$, with~$\theta$ the angle between incoming and outgoing particle.
The divergence for forward scattering~$\theta \xrightarrow{} 0$ is regulated by Debye screening.
The resulting momentum-transfer cross section is
\begin{equation}
\begin{split}
    \sigma_{DB}(v) &= 2\pi \int_{-1}^{\cos\theta_{\rm Debye}} d\cos\theta \frac{\epsilon^2 \alpha_{\rm EM}^2 (m_B+m_D)^2}{2m_B^2m_D^2 v_{DB}^4 (1-\cos\theta)} \left\{ 2 -v_{DB}^2 (1-\cos\theta) +\frac{m_B^2m_D^2}{(m_B+m_D)^4} v_{DB}^4 (1-\cos\theta)^2 \right\} \\
    &= \frac{4\pi\epsilon^2 \alpha_{\rm EM}^2 (m_B+m_D)^2}{2m_B^2m_D^2 v_{DB}^4} \left\{ \left[\log 2-\log(1-\cos\theta_{\rm Debye})\right] -v_{DB}^2 +\frac{m_B^2m_D^2}{(m_B+m_D)^4} v_{DB}^4 \right\} \\
    &\simeq \frac{4\pi\epsilon^2 \alpha_{\rm EM}^2}{\mu_{DB}^2 v_{DB}^4} \log\left(\frac{2}{\theta_{\rm Debye}}\right) \, ,
\end{split}
\end{equation}
where~$\mu_{DB}=m_Dm_B/(m_D+m_B)$, and~$\theta_{\rm Debye} \ll 1$ is the Debye angle, which in our case reads
\begin{equation}
    \theta_{\rm Debye} = \frac{m_{\rm Debye}}{\mu_{DB}v_{DB}} \, , \qquad m_{\rm Debye} = \sqrt{ 4\pi \alpha_{\rm EM} \left[ \frac{n_p+n_e}{T_b} +\frac{\epsilon^2 (n_{p_D}+n_{e_D})}{T_{b,\rm dark}} \right] } \, .
\end{equation}

The Debye screening length in a plasma can be estimated as in standard textbooks~\cite{Salzmann1998}.
Assume a structureless pointlike charge~$q_0$ located at~$r=0$, the electric potential around it is given by Poisson equation
\begin{equation}
    \nabla^2 V(\vec{r}) = -\frac{\rho}{\epsilon_0} = -\left[ en_p(\vec{r})-en_e(\vec{r}) +\epsilon e n_{p_D}(\vec{r}) -\epsilon e n_{e_D}(\vec{r}) \right] \, .
\end{equation}
An approximate solution can be found when the charged particles are moving almost freely in space and only weakly affect each other's motion, namely, at high temperature and low ion-ion interactions.
This is the requirement of a weakly-interacting (or uncorrelated, or weakly-coupled) plasma.
For an ion-electron plasma, the condition reads
\begin{equation}
    \frac{eV(\vec{r})}{T} \ll 1 \, .
\end{equation}
The corresponding requirement in our case is that~$q_XV(\vec{r})/T_X \ll 1$ for each charged particle.
In practice, in the dark sector, the decrease in temperature is compensated by the smallness of~$q_X = \epsilon e$, and the main contribution to the screening comes from the visible sector.

To first order in the argument of the exponentials,
\begin{equation}
\begin{split}
    \nabla^2 V(\vec{r}) &\simeq -\left[ e n_p\left(1-\frac{eV(\vec{r})}{T_b}\right) -e n_e\left(1+\frac{eV(\vec{r})}{T_b}\right) +\epsilon e n_{p_D}\left(1-\frac{\epsilon eV(\vec{r})}{T_{b,\rm dark}}\right) -\epsilon e n_{e_D}\left(1+\frac{\epsilon eV(\vec{r})}{T_{b,\rm dark}}\right) \right] \\
    &= e^2 V(\vec{r}) \left[ \frac{n_p+n_e}{T_b} +\frac{\epsilon^2 (n_{p_D}+n_{e_D})}{T_{b,\rm dark}} \right] \\
    &\equiv \frac{1}{D^2} V(\vec{r}) \, ,
\end{split}
\end{equation}
where the number densities are the background ones, and we have used the assumption that both visible and dark sector are on average charge neutral.
The quantity
\begin{equation}
    D = \sqrt{ \frac{1}{4\pi \alpha_{\rm EM}} \left[ \frac{n_p+n_e}{T_b} +\frac{\epsilon^2 (n_{p_D}+n_{e_D})}{T_{b,\rm dark}} \right]^{-1} }
\end{equation}
is the Debye screening length.
The Debye mass is its inverse,
\begin{equation}
    m_{\rm Debye} = \frac{1}{D} \, .
\end{equation}
The Debye angle is defined as
\begin{equation}
    \theta_{\rm Debye} = \frac{m_{\rm Debye}}{\mu_{DB} v_{DB}} \, ,
\end{equation}
with~$\mu_{DB}$ and~$v_{DB}$ being the reduced mass and the relative velocity between the two scattering particles.
We can approximate~$v_{DB}^2 \simeq T_b/m_B +T_{b,\rm dark}/m_D +V_{\rm rms}^2/3$.
In the case of DM-free electron/proton scattering, if the thermal velocity is mainly determined by the baryons then~$v_{DB}^2 \simeq T_b/m_B$, the Debye logarithm is
\begin{equation}
    \log\left(\frac{2}{\theta_{\rm Debye}}\right) = \frac{1}{2} \log\left(\frac{4 \mu_{DB}^2 v_{DB}^2 T_b}{4\pi \alpha_{\rm EM} n_e}\right) \simeq \frac{1}{2} \log\left(\frac{4 m_{e,p} T_b^2}{4\pi \alpha_{\rm EM} x_e n_{\rm H}}\right) \, .
\end{equation}

The Debye logarithm introduces a velocity dependence on top of the~$v^{-4}$ scaling, and furthermore the Debye mass evolves with redshift as roughly~$m_{\rm Debye} \propto (1+z)$, but both effects are mild~\cite{Dvorkin:2019zdi,Nguyen:2021cnb}.

\section{Tight-coupling regimes}
\label{app:TCAs}

The full evolution equations for (dark) baryon and (dark) radiation temperature are
\begin{align}
    \frac{dT_b}{dz} &= \frac{1}{1+z} \left[ 2T_b +\frac{\Gamma_{\rm C}}{H}(T_b-T_{\gamma}) +\frac{2 \RDMSM'}{H}(T_b-T_{b,\rm dark}) \right] \, , \\
    \frac{dT_{b,\rm dark}}{dz} &= \frac{1}{1+z} \left[ 2T_{b,\rm dark} +\frac{2 \left(\Pi_{\rm p-r} -\Pi_{\rm p-i} -\Pi_{\rm ff} -\Pi_{\rm R}\right)}{3k_B n_{b,\rm dark} H} +\frac{\Gamma_{\rm C,dark}}{H}(T_{b,\rm dark}-T_{\gamma,\rm dark}) +\frac{2 \RDMSMdark'}{H}(T_{b,\rm dark}-T_b) \right] \, , \\
    \frac{dT_{\gamma}}{dz} &= T_{0,\gamma} \, , \\
    \frac{dT_{\gamma,\rm dark}}{dz} &= T_{0,\gamma,\rm dark} \, ,
\end{align}
where~\cite{Cyr-Racine:2012tfp,Seager:1999km,Bansal:2022qbi}
\begin{align}
    \Gamma_{\rm C,dark} &= \frac{64 \pi^3 \alpha_D^2 T_{\gamma,\rm dark}^4}{135 m_{e_D}^3} \frac{x_{e,\rm dark}}{1+x_{e,\rm dark}} \left[ 1 + \left(\frac{m_{e_D}}{m_{p_D}}\right)^3 \right] \, , \\
    \Pi_R &\simeq \frac{430080\zeta(9)\alpha_D^2n_{{\rm H}_D}(1-x_{e,\rm dark})}{\pi^2B_D^4m_{{\rm H}_D}m_{e_D}} T_{\gamma,\rm dark}^9 r \, , \\
    \Pi_{\rm ff} &\simeq \frac{16\alpha_D^3\Bar{g}_{\rm ff}x_{e,\rm dark}^2n_{{\rm H}_D}^2\sqrt{2\pi T_{b,\rm dark}}}{(3\mu_D)^{3/2}} \left( \frac{\pi^2 r (1+2 r) -6\zeta(3) r^2}{6} \right) \, , \\
    \Pi_{\rm p-r} &= \frac{2\alpha_D^3 \sqrt{2\pi T_{b,\rm dark}}}{3 \mu_D^{3/2}} x_{e,\rm dark}^2 n_{{\rm H}_D}^2 F_{\rm p-r} \, , \\
    \Pi_{\rm p-i} &= \frac{\alpha_D^3 T_{\gamma,\rm dark}^2}{3\pi} x_{2s} n_{{\rm H}_D} e^{-B_D/(4T_{\gamma,\rm dark})} F_{\rm p-i} \, ,
\end{align}
with~$n_{{\rm H}_D}$ being the number density of dark protons, both free and bound in dark atoms, and
\begin{equation}
    r = \frac{T_{\gamma,\rm dark}-T_{b,\rm dark}}{T_{\gamma,\rm dark}} \, .
\end{equation}
We note that Ref.~\cite{Bansal:2022qbi} contains a typographical error in the denominator of the corresponding expression; the implementation in CLASS ADM is unaffected.
We define
\begin{equation}
    \Tilde{\Gamma}_{\rm eff} \equiv \Gamma_{\rm C,dark}+\Tilde{\Gamma}_{\rm R}+\Tilde{\Gamma}_{\rm ff} \, ,
\end{equation}
where the effective rates are defined from the energy exchange rates per unit time per unit volume, as
\begin{align}
    & \Tilde{\Gamma}_{\rm R} \equiv \frac{2\Pi_{\rm R}}{3 k_B n_{b,\rm dark} (T_{\gamma,\rm dark}-T_{b,\rm dark})} \, , \\
    & \Tilde{\Gamma}_{\rm ff} \equiv \frac{2\Pi_{\rm ff}}{3 k_B n_{b,\rm dark} (T_{\gamma,\rm dark}-T_{b,\rm dark})} \, .
\end{align}
Additionally, we introduce the following parameters,
\begin{equation}
    \epsilon_{\gamma b} = \frac{H}{\Gamma_{\rm C}} \, , \qquad \epsilon_{\gamma b, \rm dark} = \frac{H}{\Tilde{\Gamma}_{\rm eff}} \, .
\end{equation}

In the derivation of the tight-coupling approximation (TCA) between baryons and photons, and analogously in the dark sector, we neglect the derivatives of~$x_e$ and~$x_{e,\rm dark}$ with respect to redshift.
In the SM this is exact in the regime where the Compton TCA is applied (before HeIII recombination, as in CLASS); in the dark sector, the neglected terms would shift~$T_{b,\rm dark}$ by at most~$\mathcal{O}(\epsilon_{\gamma b, \rm dark}) \lesssim 10^{-3}$.
The following useful relations are employed
\begin{equation}
    \frac{1}{\Gamma_{\rm C,dark}}\frac{d\Gamma_{\rm C,dark}}{dz} \simeq \frac{4}{1+z} \, ,
\end{equation}
\begin{equation}
    \frac{d\Tilde{\Gamma}_{\rm eff}}{dz} \simeq \frac{4}{1+z}\Gamma_{\rm C,dark} +\frac{8}{1+z}\Tilde{\Gamma}_{\rm R} +\frac{5}{2(1+z)}\Tilde{\Gamma}_{\rm ff} \, ,
\end{equation}
\begin{equation}
    \frac{1}{\RDMSM'}\frac{d\RDMSM'}{dz} = \frac{1}{\RDMSMdark'}\frac{d\RDMSMdark'}{dz} \, ,
\end{equation}
\begin{equation}
    \frac{1}{\RDMSMdark'}\frac{d\RDMSMdark'}{dz} \simeq
\frac{\sum_{DB} W_{DB}\left[\frac{3}{1+z}
  -\frac{3}{2u_{DB}^2}\left(\frac{1}{m_B}\frac{dT_b}{dz}
  +\frac{1}{m_D}\frac{dT_{b,\rm dark}}{dz}\right)\right]}
     {\sum_{DB} W_{DB}} \, , \qquad
W_{DB} \equiv \frac{\rho_D\,\rho_B}{m_D^2\, m_B^2\, u_{DB}^3} \, ,
\end{equation}
assuming~$V_{\rm rms}^2 \simeq \text{constant}$ at early times.
\newline

\textbf{Compton TCA.}
We start with the familiar case of the SM Compton TCA.
Under the assumption that the baryon temperature closely tracks the radiation temperature,
\begin{equation}
    \epsilon_{\gamma b} = \frac{H}{\Gamma_{\rm C}} \, , \qquad \epsilon_{\gamma b} \ll 1 \, , \qquad T_b \simeq T_{\gamma} +\mathcal{O}(\epsilon_{\gamma b}) \, ,
\end{equation}
and the evolution of the temperature difference
\begin{equation}
    T_b-T_{\gamma} \simeq \epsilon_{\gamma b} \left[ -T_{\gamma} -\frac{2 \RDMSM'}{H}(T_{\gamma}-T_{b,\rm dark}) \right] +\mathcal{O}(\epsilon_{\gamma b}^2) \,
\end{equation}
is
\begin{equation}
\begin{split}
    \frac{d}{dz}(T_b-T_{\gamma}) \simeq & \frac{1}{1+z}\frac{d}{d\ln a}(\epsilon_{\gamma b}T_{\gamma}) \\
    & -\frac{2\RDMSM'}{\Gamma_{\rm C}}\left[\left(\frac{1}{\RDMSM'}\frac{d\RDMSM'}{dz}-\frac{1}{\Gamma_{\rm C}}\frac{d\Gamma_{\rm C}}{dz}\right)(T_{\gamma}-T_{b,\rm dark}) +\frac{dT_{\gamma}}{dz} -\frac{dT_{b,\rm dark}}{dz}\right] \, .
\end{split}
\end{equation}

\textbf{Dark Compton TCA.}
We review the derivation of the DM-DR TCA, which is implemented in \CLASSADM, and highlight the new terms due to DM-SM interactions.
The DM-DR TCA assumption for the dark sector is
\begin{equation}
    \epsilon_{\gamma b,\rm dark} = \frac{H}{\Tilde{\Gamma}_{\rm eff}} \, , \qquad \epsilon_{\gamma b,\rm dark} \ll 1 \, , \qquad T_{b,\rm dark} \simeq T_{\gamma,\rm dark} +\mathcal{O}(\epsilon_{\gamma b,\rm dark}) \, .
\end{equation}
The evolution of the temperature difference for the dark sector is
\begin{equation}
\begin{split}
    \frac{d}{dz}(T_{b,\rm dark}-T_{\gamma,\rm dark}) \simeq &\frac{d\epsilon_{\gamma b,\rm dark}}{dz} \left[ -T_{\gamma,\rm dark} -\frac{2\RDMSMdark'}{H}(T_{\gamma,\rm dark}-T_b) \right] \\
    & +\epsilon_{\gamma b,\rm dark} \left[ -\frac{T_{\gamma,\rm dark}}{1+z} -\frac{d}{dz}\left(\frac{2\RDMSMdark'}{H}(T_{\gamma,\rm dark}-T_b)\right) \right] +\mathcal{O}(\epsilon_{\gamma b,\rm dark}^2)
\end{split}
\end{equation}
with
\begin{equation}
    \frac{d\epsilon_{\gamma b,\rm dark}}{dz} = \frac{dH}{dz}\frac{1}{\Tilde{\Gamma}_{\rm eff}} -\frac{H}{\Tilde{\Gamma}_{\rm eff}^2}\frac{d\Tilde{\Gamma}_{\rm eff}}{dz} \, .
\end{equation}
The derivative of the effective rate, assuming~$x_{e,\rm dark}$ to be roughly constant and neglecting powers of~$(T_{\gamma,\rm dark}-T_{b,\rm dark}) \sim \mathcal{O}(\epsilon_{\gamma b,\rm dark})$, is
\begin{equation}
    \frac{d\Tilde{\Gamma}_{\rm eff}}{dz} = \frac{4}{1+z} \Gamma_{\rm C,dark} +\frac{8}{1+z} \Tilde{\Gamma}_{\rm R} +\frac{3}{1+z} \Tilde{\Gamma}_{\rm ff} -\frac{1}{2}\frac{1}{1+z} \Tilde{\Gamma}_{\rm ff} \, .
\end{equation}
Then
\begin{equation}
\begin{split}
    \frac{d(T_{b,\rm dark}-T_{\gamma,\rm dark})}{dz} \simeq & -\frac{dH}{dz}\frac{T_{\gamma,\rm dark}}{\Tilde{\Gamma}_{\rm eff}} +\frac{H T_{\gamma,\rm dark}}{\Tilde{\Gamma}_{\rm eff}^2 (1+z)} \left( 3 \Gamma_{\rm C,dark} +7 \Tilde{\Gamma}_{\rm R} +\frac{3}{2} \Tilde{\Gamma}_{\rm ff} \right) \\
    & -\frac{H}{\Tilde{\Gamma}_{\rm eff}} \frac{2\RDMSMdark'}{H} \left[ \left( \frac{1}{\RDMSMdark'} \frac{d\RDMSMdark'}{dz} -\frac{1}{\Tilde{\Gamma}_{\rm eff}} \frac{d\Tilde{\Gamma}_{\rm eff}}{dz} \right) (T_{\gamma,\rm dark}-T_b) +\frac{dT_{\gamma,\rm dark}}{dz} -\frac{dT_b}{dz} \right] \\
    & +\mathcal{O}(\epsilon_{\gamma b,\rm dark}^2) \, .
\end{split}
\end{equation}

In \CLASSADM, the dark TCA regime is switched on whenever any of the Compton, Rayleigh, or free-free rates is larger than~$2 \times 10^3$ times the Hubble rate.
\newline

In principle, a third TCA regime is possible, where all the fluids are tightly coupled together at early times and share the same temperature.
We refer to the appendix of Ref.~\cite{Flitter:2023mjj} for a discussion of this regime in the context of DM-baryon scattering.
Here, we avoid such regimes by imposing the hard prior cut on~$\ratioDMSMtoDMDR$, as described in the main text.

\section{Number density rescalings in the temperature equations}
\label{app:temperature_eqs}

Throughout this appendix we distinguish between a \emph{thermal population} and a \emph{scattering species} within it.
On the dark side,~$\chi$ denotes the set of all dark-sector particles that share a common temperature~$T_\chi$: in our case the free dark protons, the free dark electrons and the dark atoms, which are kept in kinetic equilibrium by dark Compton scattering and by collisions among themselves.
Its number density is~$n_\chi$.
A species~$D \subset \chi$ is one component of this population that actually scatters off Standard Model particles, e.g.,~the free dark protons, with number density~$n_D$ (so that~$n_{p_D} = x_D n_{{\rm H}_D}$ for free dark protons, with~$x_D$ the dark ionization fraction, and~${\rm H}_D$ denoting the dark hydrogen nuclei, both free and bound in neutral atoms).
On the visible side,~$b$ denotes the full baryonic population at temperature~$T_b$, with number density~$n_b$, and~$B \subset b$ a single target species, free protons or free electrons, with number density~$n_B$.
In the single-species case of Refs.~\cite{Dvorkin:2013cea,Boddy:2018wzy} one has~$D = \chi$ and~$B = b$, and the distinction is irrelevant.
In the atomic dark matter case, it is not, because only the ionized fraction of the dark sector scatters, while the whole dark sector shares the deposited heat.
The population quantities of this appendix map onto the main text as~$T_{\chi} = T_{b,\rm dark}$,~$n_{\chi} = n_{b,\rm dark}$, and so on. 

In the presence of DM-baryon scatterings with one dark matter species, the dark matter and baryon temperatures evolve as [e.g.,~see~Eqs.~(A21)-(A23) in Ref.~\cite{Boddy:2018wzy}]
\begin{equation}
    \Dot{T}_\chi +2\frac{\Dot{a}}{a}T_\chi = 2 R_{D}' (T_b-T_\chi) \, , \qquad \Dot{T}_b +2\frac{\Dot{a}}{a}T_b = 2\frac{\mu_b}{m_{D}} \frac{\rho_{D}}{\rho_b} R_{D}' (T_\chi-T_b) \, ,
\end{equation}
where dots denote derivatives with respect to conformal time,~$\mu_b$ is the mean molecular weight of the baryons, so that~$\mu_b \rho_D/(m_D \rho_b) = n_D/n_b$, and we neglect corrections that depend on the ratio of relative velocity to thermal velocity dispersion, i.e.,~we work in the limit~$v_{Db}/u_{Db} \ll 1$.
In the numerical implementation, the relative bulk velocity enters through its root-mean-square value, as explained in the main text,~$u_{DB}^2 \to u_{DB}^2 + V_{\rm rms}^2/3$, with~$V_{\rm rms}^2 = 10^{-8}\,c^2$ for~$z > 10^3$ and diluting with the expansion of the Universe afterwards, following Ref.~\cite{Dvorkin:2013cea}.
The expressions below are written without this term for clarity.

The heat-transfer rate is
\begin{equation}
    R_{D}' = \frac{m_{D}}{m_{D}+m_b}R_{D} \, ,
\end{equation}
and the momentum-transfer rate is~\cite{Dvorkin:2013cea,Boddy:2018wzy}
\begin{equation}
    R_{D} = \frac{a \rho_b \sigma_{0} \mathcal{N}_{-4}}{(m_{D}+m_b) u_{Db}^3} \, , \qquad \mathcal{N}_{-4}=\frac{1}{3}\sqrt{\frac{2}{\pi}} \, .
\end{equation}
In the presence of multiple target baryon species, the rates generalize to~\cite{Boddy:2022tyt,Driskell:2022pax}
\begin{equation}
\label{eq:rates_multi}
    R_{DB} = \frac{a \rho_B \sigma_{0,DB} \mathcal{N}_{-4}}{(m_{D}+m_B) u_{DB}^3} \, , \qquad R_{D} = \sum_B R_{DB} \, , \qquad R_{D}' =  \sum_B \frac{m_{D}}{m_{D}+m_B} R_{DB} \, .
\end{equation}

The temperature of a population is defined through the mean kinetic energy of \emph{all} its members, $\tfrac{3}{2}\, n_\chi T_\chi = \sum_{i \in \chi} g_i \int \tfrac{d^3 p}{(2\pi)^3}\, \tfrac{p^2}{2 m_i}\, f_i(p)$, and its evolution equation is obtained by summing the second moment of the Boltzmann equation over the species in~$\chi$ and dividing by~$n_\chi$.
Only the species~$D$ that scatter have a non-vanishing collision term with~$B$, and that collision term is proportional to~$n_D n_B$, since it involves the product~$f_D f_B$ of the two colliding distributions.
Explicitly,
\begin{equation}
\begin{split}
    \dot T_\chi + 2\frac{\dot a}{a} T_\chi
    &= \frac{1}{n_\chi} \sum_{D \subset \chi} g_D \int \frac{d^3\vec{p}}{(2\pi)^3}\,
       \frac{p^2}{3 m_D}\, \frac{a}{E}\, \mathbb{C}\!\left[f_D(p)\right] \\
    &= -\frac{a}{n_\chi} \sum_{D \subset \chi} \sum_{B \subset b} g_D g_B
       \int d\tilde p_1\, d\tilde p_2\, d\tilde p_1'\, d\tilde p_2'\,
       \frac{p_1^2}{3 m_D}\,(2\pi)^4 \delta^{(4)}(\tilde p_1+\tilde p_2-\tilde p_1'-\tilde p_2')
       \\
    & \qquad \times \left|\mathcal M_{DB}\right|^2
       \left[ f_D(p_1) f_B(p_2) - f_D(p_1') f_B(p_2') \right] .
\end{split}
\end{equation}
With Maxwell-Boltzmann distributions normalized to the densities of the scattering species,
\begin{equation}
    f_D(\vec{v}_1) = \frac{n_D}{g_D}\,
        \frac{(2\pi)^{3/2}}{m_D^3 u_D^3}\,
        \exp\!\left(-\frac{\vec{v}_1^{\,2}}{2 u_D^2}\right), \qquad
    f_B(\vec{v}_2) = \frac{n_B}{g_B}\,
        \frac{(2\pi)^{3/2}}{m_B^3 u_B^3}\,
        \exp\!\left(-\frac{\vec{v}_2^{\,2}}{2 u_B^2}\right),
\end{equation}
where~$u_D = \sqrt{T_\chi/m_D}$ and $u_B = \sqrt{T_b/m_B}$, the~$g$-factors cancel and each term in the double sum carries the prefactor~$n_D n_B / n_\chi$.
The factor~$n_B$ is what makes the rate~$R_{DB}$ proportional to~$\rho_B$, as in Eq.~\eqref{eq:rates_multi}.
The remaining factor~$n_D / n_\chi$ is the fraction of the thermal population that takes part in the scattering.
When~$D = \chi$, it equals one and the standard single-species result is recovered; in our case, it equals the dark ionization fraction weighted by the abundance of the relevant species,~$x_D/(1 + x_D)$ for each of~$D=\{p_D,e_D\}$, since no dark helium is present.
The same argument on the visible side, where the heat is shared by all of~$b$ but deposited only on~$B$, produces the factor~$n_D / n_b$ in the baryon equation.
The generalized temperature equations therefore read
\begin{equation}
    \dot T_\chi + 2\frac{\dot a}{a} T_\chi
      = 2 \sum_{D \subset \chi} \frac{n_D}{n_\chi}\, R'_{D}\,(T_b - T_\chi), \qquad
    \dot T_b + 2\frac{\dot a}{a} T_b
      = 2 \sum_{D \subset \chi} \frac{n_D}{n_b}\, R'_{D}\,(T_\chi - T_b),
\end{equation}
with~$R'_D$ defined in Eq.~\eqref{eq:rates_multi}.

\section{Energy density rescalings in the perturbation equations}
\label{app:perturbations_eqs}

We use the notation of Appendix~\ref{app:temperature_eqs}:~$D \in \{e_D, p_D\}$ is a dark species scattering off a Standard Model species~$B \in \{e, p\}$,~$\chi$ is the full dark-sector matter population (free~$e_D$, free~$p_D$ and dark atoms) and~$b$ the full baryon population. For a generic species~$s$ with mass~$m_s$, number density~$n_s$, bulk velocity~$\vec{v}_s$ and thermal dispersion~$u_s$, the Maxwell--Boltzmann distribution is
\begin{equation}
    f_s(\vec{w}) = \frac{n_s}{(2\pi)^{3/2} u_s^3}
      \exp\left\{-\frac{(\vec{w}-\vec{v}_s)^2}{2u_s^2}\right\}, \qquad
    n_s = \int d^3\vec{w} f_s(\vec{w}), \qquad
    \vec{v}_s = \frac{1}{n_s}\int d^3\vec{w} \, \vec{w} \, f_s(\vec{w}).
\end{equation}
In the non-relativistic regime the Boltzmann equation for~$f_s$ reads
\begin{equation}
    \mathbb{L}[f_s](\vec{w}) = \sum_{t} \int d^3\vec{w}_t \int d\Omega\,
      \frac{d\sigma_{st}}{d\Omega}\, g\,
      \left[ f_s(\vec{w}') f_t(\vec{w}_t') - f_s(\vec{w}) f_t(\vec{w}_t) \right],
\end{equation}
where~$\mathbb{L}$ is the Liouville operator, the sum runs over the target species~$t$, primes denote post-collision velocities and~$\vec{g} = \vec{w}_t - \vec{w}$ is the relative velocity.

\textbf{Momentum transfer.}
Multiplying by~$m_s \vec{w}$ and integrating over~$\vec{w}$ gives the momentum-density exchange rate.
Using the standard symmetries of the collision integral,
\begin{equation}
    \int d^3\vec{w}\, m_s\vec{w} \left(\frac{\partial f_s}{\partial t}\right)_{\rm coll}
    = \sum_t \int d^3\vec{w}\, d^3\vec{w}_t \int d\Omega\,
      \frac{d\sigma_{st}}{d\Omega}\, g\, m_s(\vec{w}'-\vec{w})\, f_s(\vec{w}) f_t(\vec{w}_t).
\end{equation}
In terms of the center-of-mass and relative velocities,~$\vec{G} = (m_s\vec{w} + m_t\vec{w}_t)/(m_s+m_t)$ and~$\vec{g}$, one has~$m_s(\vec{w}'-\vec{w}) = \mu_{st}(\vec{g}-\vec{g}\,')$ with~$\mu_{st} = m_s m_t/(m_s+m_t)$ the reduced mass.
Since~$|\vec{g}\,'| = |\vec{g}|$, the angular integral produces the momentum-transfer cross section,~$\int d\Omega\,(d\sigma_{st}/d\Omega)(\vec{g}-\vec{g}\,') = \vec{g}\,\sigma_{\rm MT}(g)$, with~$\sigma_{\rm MT}(g) = \sigma_{0,st}\, g^{n}$ for a power-law interaction.
The product~$f_s f_t$ factorizes into a Gaussian in~$\vec{G}$, which integrates to one, and a Gaussian in~$\vec{g}$ centred on~$\vec{v}_{st} \equiv \vec{v}_t - \vec{v}_s$ with dispersion~$u_{st}^2 = u_s^2 + u_t^2$. Hence
\begin{equation}
\begin{split}
    \int d^3\vec{w}\, m_s\vec{w} \left(\frac{\partial f_s}{\partial t}\right)_{\rm coll}
    &= \sum_t \mu_{st}\, n_s n_t\, \sigma_{0,st}
       \frac{1}{(2\pi)^{3/2} u_{st}^3}
       \int d^3\vec{g}\; g^{n+1}\,\vec{g}\;
       e^{-(\vec{g}-\vec{v}_{st})^2/(2u_{st}^2)} \\
    &= \sum_t \mu_{st}\, n_s n_t\, \sigma_{0,st}\, \mathcal N_n\,
       u_{st}^{n+1}\, \vec{v}_{st}\;
       {}_1F_1\!\left(-\frac{n+1}{2},\frac{5}{2},-\frac{r_{st}^2}{2}\right),
       \qquad r_{st} = \frac{v_{st}}{u_{st}},
\end{split}
\end{equation}
where~$\mathcal N_n = 2^{(5+n)/2}\,\Gamma(3+n/2)/(3\sqrt\pi)$~\cite{Dvorkin:2013cea}.
For~$n=-4$ and~$r_{st} \ll 1$ the hypergeometric function reduces to one and~$\mathcal N_{-4} = \tfrac13\sqrt{2/\pi}$ (see Appendix~A of Ref.~\cite{Boddy:2018wzy}; the overall sign differs from there because we define~$\vec{g}$ as target minus projectile).
Since the left-hand side is the collisional part of~$\partial(\rho_s\vec{v}_s)/\partial t$, converting to conformal time gives
\begin{equation}
    \left.\frac{\partial(\rho_s\vec{v}_s)}{\partial\tau}\right|_{\rm coll}
      = \sum_t R_{st}\, \rho_s\, (\vec{v}_t - \vec{v}_s), \qquad
    R_{st} \equiv \frac{a\, \rho_t\, \sigma_{0,st}\, \mathcal N_{-4}}
                       {(m_s+m_t)\, u_{st}^3},
\label{eq:app_momentum_rate}
\end{equation}
which is the momentum-transfer rate of App.~\ref{app:temperature_eqs}.
Note that~$R_{st}\rho_s = R_{ts}\rho_t$, so that momentum is conserved between the two populations.

\paragraph{Dark sector.}
We assume that only the free dark species scatters with free SM particles, and we assume that free dark electrons, free dark protons and dark atoms are tightly coupled among themselves so that they share a common bulk velocity~$\vec{v}_\chi$, while all SM species share~$\vec{v}_b$.
Adding the momentum equations of the three dark components, using~$\rho_\chi = \rho_{e_D,\rm free}+\rho_{p_D,\rm free}+\rho_{{\rm H}_D,\rm bound} \propto a^{-3}$ to pull the total density out of the time derivative, and applying Eq.~\eqref{eq:app_momentum_rate} to~$s = D$ and~$t = B$,
\begin{equation}
    \frac{\partial\vec{v}_\chi}{\partial\tau} + \dots
      = \sum_{D}\frac{\rho_D}{\rho_\chi}\sum_B R_{DB}\;(\vec{v}_b-\vec{v}_\chi)
      = \left[\frac{\rho_{e_D}}{\rho_\chi}\left(R_{e_D p}+R_{e_D e}\right)
             +\frac{\rho_{p_D}}{\rho_\chi}\left(R_{p_D p}+R_{p_D e}\right)\right]
        (\vec{v}_b-\vec{v}_\chi),
\end{equation}
which defines the energy-density weights used in the main text.

\paragraph{Baryons.}
The same steps for~$s = B$ and~$t = D$, together with~$R_{BD}\rho_B = R_{DB}\rho_D$, give
\begin{equation}
    \frac{\partial\vec{v}_b}{\partial\tau} + \dots
      = -\sum_{D}\frac{\rho_D}{\rho_b}\sum_B R_{DB}\;(\vec{v}_b-\vec{v}_\chi)
      = -\left[\frac{\rho_{e_D}}{\rho_b}\left(R_{e_D p}+R_{e_D e}\right)
              +\frac{\rho_{p_D}}{\rho_b}\left(R_{p_D p}+R_{p_D e}\right)\right]
         (\vec{v}_b-\vec{v}_\chi),
\end{equation}
where~$\rho_b$ is the total baryon density, including helium and neutral atoms, since the whole baryon fluid shares~$\vec{v}_b$.

\section{Posteriors}

\begin{figure}
    \centering
    \includegraphics[width=0.49\linewidth]{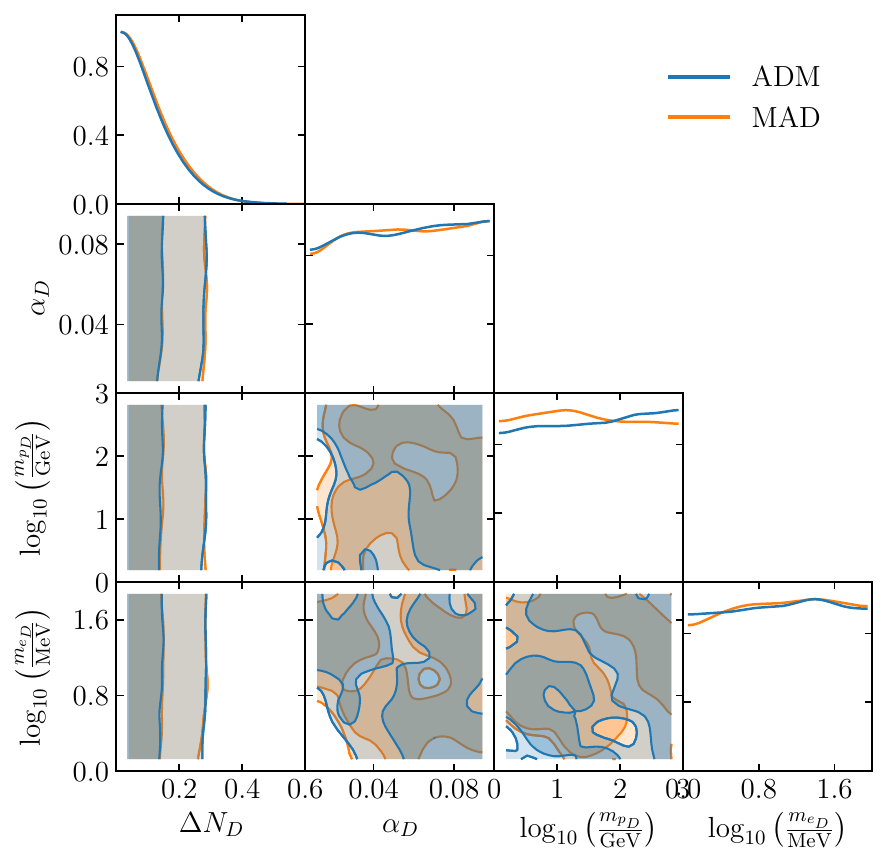}
    \caption{Two-parameter projections of the posterior distribution for the ADM parameters~$\{ \Delta N_D, \, \alpha_D, \, m_{p_D}, \, m_{e_D} \}$, for MAD and ADM in orange and blue respectively.
    Dark and light contours correspond to~$68\%$ and~$95\%$ confidence level, respectively.}
    \label{fig:tiles_ADMandMAD}
\end{figure}

Figure~\ref{fig:tiles_ADMandMAD} shows two-parameter projections of the posterior for the ADM parameters,~$\left\{ \Delta N_D, \, \alpha_D, \, m_{p_D}, \, m_{e_D} \right\}$, after marginalizing over the rest.
Two main results emerge.
First, as noted in the main text and in agreement with the findings of Ref.~\cite{Bansal:2022qbi}, except for the strong bound on the DR contribution to~$\Delta N_{\rm eff}$, other combinations of two parameters remain nearly unconstrained.
Second, incorporating millicharge does not improve nor spoil this picture.

\begin{figure}
    \centering
    \includegraphics[width=0.85\linewidth]{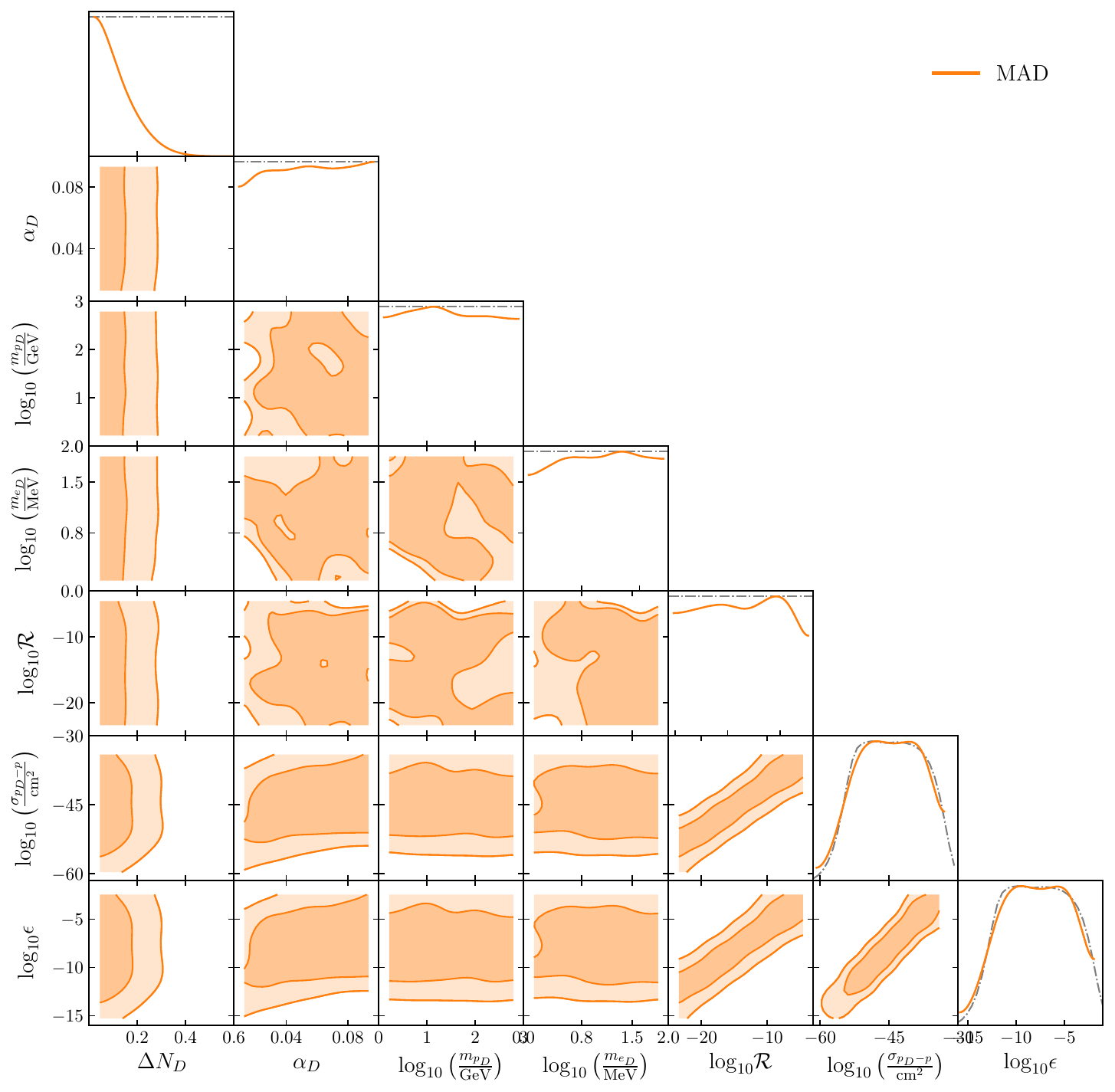}
    \caption{Two-parameter projections of the posterior distribution for MAD parameters, both those varied in the MCMC~$\{ \Delta N_D, \, \alpha_D, \, m_{p_D}, \, m_{e_D}, \, \ratioDMSMtoDMDR \}$ and derived ones~$\{ \sigma_{p_D-p}, \, \epsilon \}$.
    Dark and light contours correspond to~$68\%$ and~$95\%$ confidence level, respectively.
    The priors are shown in dashed gray lines.
    The MCMC scan is cut off at the upper bound~$\mathcal{R} < 10^{-3}$ imposed to preserve dark recombination; within this requirement, the corresponding derived cross sections and millicharges span the entire prior range.}
    \label{fig:tiles_fullMAD_priors}
\end{figure}

Figure~\ref{fig:tiles_fullMAD_priors} shows two-parameter projections for the MAD scenario, including the MCMC input parameters with the respective priors,~$\{ \Delta N_D, \, \alpha_D, \, m_{p_D}, \, m_{e_D}, \, \ratioDMSMtoDMDR \}$, and the derived parameters~$\{ \sigma_{p_D-p}, \, \epsilon \}$.
The shape of the posteriors at high values of~$\ratioDMSMtoDMDR$ is insensitive to the (unphysical) lower bound of the prior, but it is instead affected by the (physical) upper cutoff~$\ratioDMSMtoDMDR = 10^{-3}$.
The upper limit is chosen to ensure that the process of dark recombination is not spoiled by DM-SM interactions.
Within this regime, all the values of cross section and corresponding millicharge spanned by the prior are still allowed, after marginalizing over the rest of the parameters.
Our aim is to establish that MAD admits a cosmologically viable region of parameter space realizing a small relic millicharged fraction, rather than to map the model’s full parameter space.
We therefore restrict the analysis to the regime in which the DM-DR tight-coupling approximation holds at dark recombination, and we use a log-uniform prior on~$\ratioDMSMtoDMDR$ to ensure adequate sampling of the cosmologically interesting cross-section range.
Other regimes are left to future work.

\section{Code}
\label{app:code}

Our code, \CLASSMAD, makes several changes to \CLASSADM, and in particular the DM-SM interactions were implemented following the formalism of Refs.~\cite{Nguyen:2021cnb,Gluscevic:2017ywp,Boddy:2018kfv} and of Ref.~\cite{Archidiacono:2019wdp}.
To include DM-SM interactions, we merge the computation of the thermodynamics module for the visible and dark sectors, which were kept separate in \CLASSADM.
Below we give more details on the input parameters and on the modifications to the thermodynamics and perturbations modules.
\newline

\CLASSMAD takes five input parameters to describe the dark sector:~$\{ \Delta N_D, \, \alpha_D, \, m_{p_D}, \, m_{e_D} \}$ to describe the ADM, and one additional parameter to characterize the interactions.
The present-day temperature of the DR is derived as~$T_{0,\rm dark} = T_{0,\rm CMB} \left( \Delta N_D / 4.4 \right)^{1/4}$.
The dark coupling constant can be passed as either~$\{ \alpha_D, \, {\rm log}_{10} \alpha_D \}$.
The masses are given as either one of~$\{ m_{p_D} \, [{\rm GeV}], \, {\rm log}_{10} \left(m_{p_D}/{\rm GeV}\right) \}$ and either one of~$\{ m_{e_D} \, [{\rm GeV}], \, {\rm log}_{10} \left(m_{e_D}/{\rm GeV}\right), \, {\rm log}_{10} \left( m_{e_D}/m_{p_D} \right) \}$.
In principle, the \CLASSADM code also allows the user to set the fraction of ADM with respect to the total dark matter energy density; however, in this work we require that ADM comprises all of the dark matter.

The DM-SM interactions can be characterized through either one of~$\{ \sigma_{0,p_D-p} \, [{\rm cm}^2], \, \log_{10} \left( \sigma_{0,p_D-p} / {\rm cm}^2 \right), \, \log_{10} \ratioDMSMtoDMDR \}$.
For debugging purposes, four flags are implemented to switch on and off the individual interaction rates between baryonic and fermionic species---this can be useful for comparison with models of interacting dark matter e.g.,~with electrons or with protons only.
The conversion between the overall rescaling of the cross sections~$\sigma_{0,p_D-p}$ and the relative importance of DM-SM to DM-DR interactions at dark recombination~$\ratioDMSMtoDMDR$ is carried out at the estimated redshift of dark decoupling~$z_{\rm dec, dark}^{\rm (estimate)}$, as detailed in the main text.

Within the code, we carry out the estimate of the millicharge parameter~$\epsilon$ from~$\sigma_{0,DB}$ by expanding Eq.~\eqref{eq:momentum_transfer_cross_section} for small~$\epsilon$, as
\begin{equation}
\label{eq:link_cross_millicharge}
    \epsilon^2 \simeq \frac{1}{2} \frac{n_p+n_e}{n_{p_D}+n_{e_D}} \frac{T_{b,\rm dark}}{T_b} \left\{ {\rm log}\left(\frac{\mu_{DB}^2v_{DB}^2T_b}{\pi \alpha_{\rm EM}(n_p+n_e)}\right) - \sqrt{ \left({\rm log}\left[\frac{\mu_{DB}^2v_{DB}^2T_b}{\pi \alpha_{\rm EM}(n_p+n_e)}\right] \right)^2 - \frac{2 \sigma_{0,DB} \mu_{DB}^2}{\pi \alpha_{\rm EM}^2} \frac{n_{p_D}+n_{e_D}}{n_p+n_e} \frac{T_b}{T_{b,\rm dark}} } \right\} \, ,
\end{equation}
where the negative sign solution is kept.
All quantities are evaluated at~$z=0$, in order to work with well-defined asymptotic quantities for the dark sector, and setting~$\{D,B\}=\{e_D,e\}$ as a reference.
Choosing a different redshift, e.g.,~around CMB times, or a different scattering process would only change the inferred value by~$\lesssim 10\%$.

We refer the reader to Ref.~\cite{Bansal:2022qbi} for an in-depth description of \CLASSADM.
Here we comment on the changes made in \CLASSMAD in the thermodynamics and perturbations modules.
\newline

The Python wrapper was modified to access some quantities of interest.
\begin{itemize}
    \item \lstinline{cross_idm_b} returns~$\sigma_{0,p_D-p} \, [{\rm cm}^2]$.
    \item \lstinline{ratio_idm_b} returns~$\ratioDMSMtoDMDR$.
    \item \lstinline{epsilon_millicharge} returns the value of the millicharge fraction~$\epsilon$ corresponding to the input parameters, Eq.~\eqref{eq:link_cross_millicharge}.
    \item \lstinline{z_d_twin_estimate} returns the estimate of the redshift of dark decoupling, obtained in \CLASSADM as~$\Gamma_{\rm C,dark}(z_{\rm dec, dark}^{\rm (estimate)}) = 100 H(z_{\rm dec, dark}^{\rm (estimate)})$, solved by bisection with the Saha ionization fraction.
    This is an estimate of the thermal decoupling redshift.
    \item \lstinline{z_d_twin} returns the dark drag redshift, corresponding to the conformal time at which the optical depth of the DM-DR plasma~$\tau_{d,\rm dark}$ crosses unity.
    This is an estimate of the kinetic decoupling redshift.
    \item \lstinline{x_e_twin_asympt} returns the value of the dark ionized fraction at late times~$x_{e,\rm dark}(z=0)$.
\end{itemize}

In the thermodynamics module, the computations of the visible and dark thermal histories have been merged.
In particular, the functions which were separate in \CLASSADM and have been merged in \CLASSMAD are
\begin{itemize}
    \item \lstinline{thermodynamics_vector_init} and \lstinline{thermodynamics_vector_init_twin},
    \item \lstinline{thermodynamics_derivs} and \lstinline{thermodynamics_derivs_twin},
    \item \lstinline{thermodynamics_source} and \lstinline{thermodynamics_source_twin}.
\end{itemize}
The differential equation solver is called on the merged list of redshift intervals, keeping track of the current approximation regime for each sector, and of which sector switched approximation latest for initialization purposes.

The set of differential equations for the evolution of the temperatures has been modified according to the approximation regime.
The evolution equations in the tight-coupling regime between baryons and radiation, and between dark baryons and dark radiation, are described in Appendix~\ref{app:TCAs}.

In the perturbations module, we closely follow the implementation of DM-baryon scattering as in Refs.~\cite{Nguyen:2021cnb,Gluscevic:2017ywp,Boddy:2018kfv} and Refs.~\cite{Archidiacono:2019wdp}, by appropriately replacing the DM-baryon scattering rate with our interaction rates, weighted by the energy densities of the interacting species, as described in Appendix~\ref{app:perturbations_eqs}.